\documentclass[10.5pt,preprint]{aastex}
\usepackage{epsfig} 
\usepackage{color}

\newcommand\etal{et~al.} 

\slugcomment{Accepted to \apj}
\shorttitle{Globular Cluster Color--Metallicity Nonlinearity: M84 in HST/WFC3 $u$-band}
\shortauthors{Yoon et al.}

\begin{document}

%\title{Nonlinear Color--Metallicity Relations of Globular Clusters.\\ IV. Testing the Nonlinearity Scenario for Color Bimodality via HST/WFC3 {\lowercase $u$}-band Photometry of M84 (NGC 4374)}
\title{Nonlinear Color--Metallicity Relations of Globular Clusters.\\ IV. Testing the Nonlinearity Scenario for Color Bimodality via {\it HST}/WFC3 $u$-band Photometry of M84 (NGC 4374)}

\author{Suk-Jin Yoon\altaffilmark{1}, 
        Sangmo T. Sohn\altaffilmark{2}, 
        Hak-Sub Kim\altaffilmark{1}, 
        Chul Chung\altaffilmark{1}, 
        Jaeil Cho\altaffilmark{1}, 
        Sang-Yoon Lee\altaffilmark{1}, 
        and John P. Blakeslee\altaffilmark{3}}
\altaffiltext{1}{Department of Astronomy and Center for Galaxy Evolution Research, 
                 Yonsei University, Seoul 120-749, Republic of Korea}
\email{sjyoon@galaxy.yonsei.ac.kr}
\altaffiltext{2}{Space Telescope Science Institute (STScI), 
                 3700 San Martin Drive, Baltimore, MD 21218}
\altaffiltext{3}{Herzberg Institute of Astrophysics, National Research Council of Canada, 
                 Victoria, BC V9E 2E7, Canada} 

%=========================================================================
%=========================================================================
\begin{abstract}
Color distributions of globular clusters (GCs) in most massive galaxies 
are bimodal. Assuming linear color-to-metallicity conversions, 
bimodality is viewed as the presence of merely two GC subsystems 
with distinct metallicities, which serves as a critical backbone of various 
galaxy formation theories. Recent studies, however, revealed that the 
color--metallicity relations (CMRs) often used to derive GC 
metallicities (e.g., CMRs of $g-z$, $V-I$ and $C-T_1$) are in fact
{\it inflected}. Such inflection can create bimodal color 
distributions if the underlying GC metallicity spread is simply broad
as expected from the hierarchical merging paradigm of galaxy formation. 
In order to test the nonlinear-CMR scenario for GC color bimodality, 
the $u$-band photometry is proposed because the $u$-related CMRs 
(e.g., CMRs of $u-g$ and $u-z$) are theoretically predicted to be 
least inflected and most distinctive among commonly used optical
CMRs. Here, we present {\it Hubble Space Telescope} ({\it HST\,})/WFC3 
$F336W$ ($u$-band) photometry of the GC system in M84, a giant 
elliptical in the Virgo galaxy cluster. 
Combining the $u$ data with the existing {\it HST} ACS/WFC $g$ and $z$ data, 
we find that the $u-z$ and $u-g$ color distributions are different from the $g-z$ distribution 
in a very systematic manner and remarkably consistent with our model 
predictions based on the nonlinear-CMR hypothesis. 
The results lend further confidence 
to validity of the nonlinear-CMR scenario as an explanation for GC color bimodality.
There are some GC systems showing bimodal spectroscopic metallicity,
and in such systems the inflected CMRs often create stronger bimodality in the color domain.
\end{abstract}

\normalsize

%=========================================================================
\section{INTRODUCTION}

\subsection{Color Bimodality of Globular Clusters}

Globular clusters (GCs) are present in galaxies of all morphological types
and contain rich information about old stellar populations.
Since GC formation occurs with starbursts in galaxies, 
they can be used to place stringent constraints on
the histories of star formation, chemical enrichment and mass assembly of 
their parent galaxies. Compared to integrated light from multiple, complex 
stellar populations of galaxies, GCs are easier to interpret thanks to 
their small internal dispersion in age and chemical abundance. Systematic 
studies on GC systems, therefore, are a powerful means of 
investigating galaxy formation and evolution
\citep[For reviews, see][]{harris91,west04,brodie06}.

One of the most remarkable developments in the field of extragalactic 
GCs over the past couple of decades is the discovery of ``bimodal'' 
distributions of GC optical colors (e.g., $C-T_{1}$, $V-I$ and $g-z$).
Ever since the first recognition and statistical study by 
\citet{zepf93}, color bimodality has been found to be a common 
feature among GC systems of the majority of massive galaxies 
\citep[e.g.,][]{ostrov93,whitmore95,mglee98,gebhardt99,
harris01,kundu01,larsen01,peng04a,peng04b,peng06,harris06,mglee08,
jordan09,sinnott10,liu11,forbes11,faifer11,foster11,chies12,forte12,young12,blom12}. 
By adopting simple linear color-to-metallicity conversions, the bimodality observed in GC color 
distributions has been generally taken as bimodal metallicity 
distributions and hence interpreted as the presence of two distinct 
GC subsystems in each galaxy. The origin of merely two GC subgroups 
within individual galaxies and its implications in the context of galaxy 
formation have attracted much interest. Scenarios have been 
put forward for the GC and galaxy formation through major mergers 
\citep{ashman92}, multiphase dissipational collapses \citep{forbes97}, 
accretion \citep{cote98}, and a hybrid of them 
\citep{mglee10,arnold11,strader11,forbes11,romanowsky12}.

\subsection{Color Bimodality and Metallicity--Color Nonlinearity}

The key assumption behind the notion that bimodal color histograms of 
GCs correspond to bimodal metallicity distributions is that optical 
colors are simple, linear proxies for metallicity. To first order, 
this is a reasonable assumption given that the mean color of bright 
giant-branch stars (i.e., the main sources of integrated optical light 
of GCs) is a strong function of metallicity. However, in order to 
examine the detailed structure of the underlying GC metallicity 
distribution functions (MDFs), one needs a more exact form of the color-metallicity 
relations (CMRs) to higher order. An oversimplified color-to-metallicity conversion may lead to falsely 
derived MDFs, which in turn would exert an adverse effect on 
interpreting the chemical evolution of GC systems and their host galaxies.

The possibility of nonlinear CMRs has been proposed and investigated 
by several studies. On observational grounds, Kissler-Patig \etal\ 
(1998) pointed out that the slope in the $V-I$ CMR becomes flatter 
toward redder colors. Peng \etal\ (2006) presented an empirical 
$g-z$ CMR that is steep for lower metallicities and shallow at higher 
metallicities. Richtler (2006) showed that scatter in CMRs 
can make a unimodal MDF appear bimodal.
On theoretical grounds, Yoon \etal\ (2006, hereafter 
Paper I) introduced wavy, nonlinear CMRs based on their stellar 
population simulations and showed that such CMRs reproduce the 
observed CMRs better than the simple, linear relations. The physical 
basis of the wavy form of their theoretical CMRs lies in the 
nonlinear metallicity dependence of the mean colors of {\it both} 
the red-giant branches and the horizontal branches (HBs) in old stellar 
populations.

Perhaps the most important implication of the inflected CMRs is that 
they can create bimodal color distributions from a unimodal 
metallicity spread through the metallicity-to-color ``projection 
effect'' (Paper I). Cantiello \& Blakeslee (2007) confirmed that 
nonlinear CMRs can produce bimodal color distributions by performing 
simulations using various stellar population models. Another 
important implication of the nonlinear CMRs is the possibility of 
deriving MDFs from existing color distributions. Yoon \etal\ (2011b, 
hereafter Paper III) converted the observed color distributions of 
GCs in several galaxies into MDFs using their theoretical CMRs, and 
compared with the observed stellar MDFs of nearby early-type galaxies. 
The GC MDFs derived this way are found to be remarkably similar to 
those of field stars in the galactic halos, implying that GC systems and their 
parent galaxies have shared a more common origin than previously 
thought.

\subsection{Testing the Metallicity--Color Nonlinearity Explanation}

Despite the broad implication of the metallicity--color nonlinearity explanation 
for GC color bimodality, still controversial are how strongly CMRs are inflected 
and how important the role of nonlinearity is in producing color bimodality. 
Obviously, the most direct way to test the veracity of the theoretical CMRs is 
establishing the empirical CMRs using spectroscopic metallicities for large sample of GCs. 
However, spectroscopic data of sufficient 
quality still remain observationally expensive. 
Despite the groundbreaking nature of the spectroscopy of extragalactic GC systems 
at distances of the Virgo cluster and beyond, the results tend to be 
limited by the sample sizes ($<$\,1\,\% of the whole GC population),
and even the best samples for observational color--metallicity calibrations 
still have significant observational scatter.
Furthermore, the absorption line strengths versus metallicity relations 
are not as simple as it might seem.  
\citet{chung13}, \citet[][hereafter Paper V]{skim13}, and S.-J. Yoon et al. (2013, in preparation)
show that the nonlinear metallicity dependence of the mean temperature
of {\it both} the red-giant branches and the HBs 
exert an appreciable effect on absorption line strengths: 
both metallicity-sensitive lines (e.g., Mg, Fe, and CaT) and Balmer lines. 
The effect brings about the GC index--metallicity relations that are nonlinear. 
Considering such metallicity--index nonlinearity is critically important 
for deriving spectroscopic metallicities accurately from various indices
and thus for establishing the correct empirical CMRs.

The metallicity--index nonlinearity issue is 
closely analogous to that of metallicity--color nonlinearity (Paper I),
and is intimately connected to obtaining spectroscopic metallicity distributions 
of GCs---the second most direct test of the nonlinear-CMR scenario for color bimodality.
High quality datasets are now becoming available for relatively nearby GC systems 
(e.g., Beasley \etal\ 2008; Woodley \& Harris 2011; Caldwell \etal\ 2009, 2011; 
Foster \etal\ 2011; Alves-Brito \etal\ 2011; Brodie \etal\ 2012; Usher \etal\ 2012; Park \etal\ 2012). 
Using Caldwell et al.'s (2011) spectroscopy on M31 GCs with unprecedented
precision and the theoretical index--metallicity relations,  
Paper V demonstrates that 
the metallicity--index nonlinearity is critical to explain 
the intriguing bimodality in index distributions of GCs in massive galaxies.
For instance, even for a unimodal underlying MDF, 
the index distribution of metallicity-sensitive Mg\,$b$ can be bimodal. 
This has been directly interpreted as a bimodal metallicity distribution, 
not considering index--metallicity nonlinearity. 
Similarly, and perhaps more importantly, 
Balmer lines (H$\beta$, H$\gamma$ and H$\delta$)
show highly inflected metallicity--index relations
and in turn exhibit very strong bimodal index distributions, 
which are routinely translated into bimodal metallicity distributions. 
Balmer lines seem to have a significant role in establishing 
the notion of bimodality in spectroscopic metallicity, 
given that most studies so far have derived spectroscopic [Fe/H]
based {\it jointly} on metal-lines and Balmer lines.

A highly complementary to spectroscopy, and more observationally efficient, method
relies on multiband photometric colors. 
Yoon \etal\ (2011a, hereafter Paper II) proposed that multiple color distributions 
allow for an important test of the color--metallicity nonlinearity scenario by Paper I. 
In essence, this technique exploits the following two facts: 
(1) if the MDF of a given GC system is truly bimodal in nature, 
any color distribution should exhibit bimodality; 
and (2) if one color distribution has a significantly different ``shape'' 
from another for the identical sample of GCs, 
the assumption that colors are linear proxies of metallicities is invalidated. 
The technique can, in principle, work for any combination of colors. 
However, it is clearly best to use the color combinations 
that provide the most contrasting case. 
Paper II has experimented with several cases and 
found that the colors based on the $u$-band are favorable. 
This is because the CMRs for the $u$-band colors 
(e.g., CMRs of $u-g$ and $u-z$) are substantially less inflected 
than those for the other commonly-used optical colors
(e.g., CMRs of $g-z$, $V-I$ and $C-T_1$).

Figure 1, taken from Paper II, demonstrates briefly how this technique works. 
This specific experiment targeted the M87 GC system, 
one of the very few early-type galaxies with existing deep $u$-band data. 
A given metallicity spread (vertical histograms in the top panels) 
is projected via the theoretical CMRs (solid lines in the top panels) 
to create the simulated color distributions (gray histograms in the middle panels). 
For the set of colors shown in the figure, the modeled distributions are 
noticeably different, especially between $g-z$ (leftmost column) and $u-g$ (rightmost): 
While the $g-z$ distribution clearly shows a bimodal case 
with roughly the same heights for the blue and red peaks, 
the bimodality in the $u-g$ distribution is substantially diminished 
with the blue peak now dominating the overall distribution. 
Remarkably, the simulated color distributions are similar to 
the observed distributions (histograms in the bottom panels), 
thereby supporting the idea of the nonlinear CMRs. 
In the same vein, 
\citet{cantiello07}, \citet{kundu07}, \citet{spitler08}, and \citet{chies11,chies12} 
highlighted the usefulness of optical--IR colors in constraining the underlying MDFs. 
More recently, \citet{blakeslee12} showed that while the color distribution 
of GCs in NGC 1399 was clearly bimodal in the optical $g-I$ color, 
the same bimodality was not present in the optical--IR $I-H$ color. 
Moreover, the color--color relation between $g-I$ and $I-H$ colors 
in this galaxy was distinctly nonlinear, 
indicating significant nonlinearity in the CMR for at least one of these colors.

In this paper, we perform the {\it HST}/WFC3 archival $u$-band photometry 
for the M84 (NGC 4374) GC system  
and apply the $u$-band technique to the system. 
M84 is a giant elliptical galaxy located in the Virgo galaxy cluster, 
and exhibits clear color bimodality in $g-z$ \citep{peng06,jordan09}. 
Section 2 presents our data reduction and photometry procedure
and describes the observational data on the GC system of M84. 
Section 3 gives the result of our simulations and compares it with the observation. 
The model based on our nonlinear-CMR scenario shows that 
the $u$-band color distributions are significantly less bimodal than that of $g-z$,
and agree well with the observation. 
Our theoretical prediction for the metallicity distribution of M84 GC system is also presented. 
Section 4 discusses the implications of our results 
for the nonlinear-CMR scenario of GC color bimodality (Section 4.1)
and for the formation of M84 (Section 4.2) in comparison with M87 (Paper II).

%=========================================================================
\section{THE M84 GLOBULAR CLUSTER SYSTEM: OBSERVATION}

\subsection{The {\it HST}/WFC3 $u$-band Imaging}

The first galaxy to which we apply the $u$-band technique proposed by 
Paper II has been selected by the procedure below. First, we considered 
galaxies that have been observed as part of the ACS Virgo Cluster 
Survey (ACSVCS; C\^ot\'e \etal\ 2004) and ACS Fornax Cluster Survey 
(ACSFCS; Jord\'{a}n \etal\ 2007). We inspected the $g-z$ color 
distributions of their GC systems and selected the galaxies showing 
clear color bimodality. Then, among them, we searched for galaxies with 
deep {\it HST} $u$-band photometry. We identified M84 (NGC 4374), a giant 
galaxy in the Virgo galaxy cluster, whose GC system exhibits clear 
bimodality in the ACS $g-z$ distribution and is one of few elliptical 
galaxies with deep WFC3/UVIS F336W images. The M84 images were 
observed as part of the science program {\it HST} GO-11583 (PI: J. 
Bregman) to constrain the star formation rate in nearby elliptical 
galaxies.

Despite the great opportunity that the $u$-band provides, observations 
of extragalactic GCs in this wavelength are lacking because the 
atmospheric transmission at $\lambda$ $<$ 4000\,{\small \AA} is 
limited for ground-based observations \citep[e.g.,][]{hkim13} and because the pre-WFC3 
detectors of {\it HST} were not efficient (WFPC2: low sensitivity; 
ACS/HRC: small field size) for systematic studies on extragalactic GCs. 
Whereas the M87 $u$-band data shown in Figure 1 required 12 orbits of 
exposure using {\it HST} WFPC2 (Paper II), the M84 WFC3 data surpass 
the WFPC2 data quality with only a fraction of exposure time thanks to 
the excellent blue sensitivity of WFC3/UVIS. Our M84 result shows that 
with two orbits of exposure, the $u-g$ color errors become as small as 
$\sim$\,0.04 mag for a typical GC with $u$ = 25. The M84 fields 
overlap with the existing ACS/WFC $g$ and $z$ observations in ACSVCS, 
yet the field of view of WFC3/UVIS is 64\,\% the size of ACS/WFC. The 
radial number density profile of GCs in M84 makes $\sim$\,85\,\% of 
the ACS/WFC GCs placed in the WFC/UVIS field of view.

\subsection{Data Reduction}

The data reduction and photometry procedure are outlined as follows.
We retrieved two drizzled WFC3 F336W images of M84 from the {\it HST} 
archive. These images are rotated with respect to each other by a 
position angle of $\sim$\,120 deg while sharing the same image centers. 
Integration time for each image is 2400 sec. Sources were detected 
using the {\it daofind} task in IRAF\footnote{IRAF is distributed by 
the National Optical Astronomy Observatory, which is operated by the 
Association of Universities for Research in Astronomy (AURA) under 
cooperative agreement with the National Science Foundation.} and 
matched with those in the catalog of GC candidates of the ACSVCS 
\citep{jordan09}. Using {\it daophot} task in IRAF, aperture 
photometry was carried out in a 3 pixel radius aperture and adjusted 
to 10 pixels using empirical aperture corrections derived via several 
bright and isolated sources in the images. These magnitudes were then 
corrected to an infinite aperture using the value of $-0.131$ which 
was derived using the encircled energy fraction provided by the WFC3 
Instrument Handbook \citep{dressel11}. Finally, we transformed the 
magnitudes to the AB system using zero points from \citet{dressel11}. 
The corrections for foreground extinction were applied following the 
same method as in \citet{jordan04}. The GC candidate catalog of the 
ACSVCS \citep{jordan09} contains bona fide GCs selected by their 
magnitudes, $g-z$ colors, and sizes. We further employed color cuts 
($g-z$\,$<$\,0.5 and $u-g$\,$<$\,0.8) to filter out potentially 
contaminating sources such as background star-forming galaxies.

The resultant M84 GCs are presented in Figure 2. Table 1 gives the ID, 
R.A., Decl., $u$-, $g$-, and $z$-band mags, and their observational 
errors for M84 GCs. In this study, we consider selected 362 GCs 
($\sigma_u$\,$<$\,0.2 mag) that have reliable $u$, $g$ and $z$ 
photometry in common, and the sample is $u$-band limited. The top 
panels of the figure are the color--magnitude diagrams and the bottom
panels are the color--color diagrams along with their color 
distributions shown by gray histograms at the top and side. In the 
color--magnitude diagrams, the split between two vertical bands of GCs
is readily visible for $g-z$, whereas it is less clear for $u-z$ and 
$u-g$. In the color--color diagrams, the red lines are our theoretical 
predictions (Tables 2 and 3) for 13-Gyr GCs from the lowest 
metallicity ([Fe/H] = $-2.5$, top left point) to the highest 
([Fe/H] = +0.5, bottom right). The crosses on each model line mark 
the uniform [Fe/H] intervals of $\Delta$[Fe/H] = 0.2 dex. The larger 
color spaces at the midpoint of, for example, the $g-z$ versus $u-z$ 
relation lead to the observed scarcity at the center of the 
corresponding color distributions.

%======================================================================
\section{THE M84 GLOBULAR CLUSTER SYSTEM: SIMULATIONS}

\subsection{Theoretical Color--Metallicity Relations Associated with
WFC3 $u_{\rm F336W}$ and ACS $g_{\rm F475W}$ and $z_{\rm F850LP}$}

Figure 3 displays the synthetic color--magnitude diagrams 
for individual stars of the model GCs and the resulting CMRs 
from the Yonsei Evolutionary Population Synthesis (YEPS) 
model\footnote{http://web.yonsei.ac.kr/cosmic/data/YEPS.htm} (Chung et al. 2013). 
This figure is similar to Figure 1 of Paper II on M87, 
but specialized for the M84 case. 
The present model for M84 differs from the Paper II model for M87 
in that this study uses the WFC3 $u_{\rm F336W}$ filter (see Figure 4) 
and the 13-Gyr model GCs. 
The upper-left quadrant shows 
the synthetic Log {$T_{\rm eff}$} versus Log $L/L_{\sun}$ diagrams, 
from which the model CMRs for $g-z$, $u-z$, and $u-g$ are generated.

The rest three quadrants of Figure 3 present the theoretical CMRs 
along with the observations. 
The upper-right quadrant shows that the $g-z$ CMR is 
an inverted S-shaped ÒwavyÓ curve, consistent with the observations.
The metal-rich ([Fe/H] $\gtrsim$ 0.0) GCs, however, show 
a $\sim$\,0.1 mag offset in $g-z$ with respect to the theoretical relation. 
Interestingly, the $g-z$ peak color of red GCs in Figure 1 (and Figure 6 in the next Section 3.2) 
is redder than the observation by the similar amount. 
The offset can be explained if metal-rich GCs are slightly younger than blue ones (by $\sim$\,2 Gyr) 
within the current uncertainty in GC age dating (e.g., Strader et al. 2005)
or if they have an extended blue HBs, as observed in the Galactic counterparts 
(e.g., NGC 6388 and NGC 6441, Rich et al. 1997; Caloi \& D'Atona 2007; Yoon et al. 2008).
On the other hand, the lower-left and right quadrants show that
the CMRs for the $u$-band colors are substantially less inflected 
than the $g-z$ CMR for the given age, 
and reproduce the observational data well.
To quantify the degree of agreement 
between the observed data and the theoretical predictions, 
we obtained the error-weighted $\chi^2$ between them: 
The reduced $\chi^2$ values are as low as 0.737, 0.948, and 0.688 
for $g-z$, $u-z$, and $u-g$, respectively.
It is interesting to notice that 
the degree of nonlinearity is in order of $g-z$, $u-z$, and $u-g$.

The present model for M84 uses {\it HST}/WFC3 $u_{\rm F336W}$ filter 
for which M84 {\it HST} images are available, 
whereas the M87 model in Paper II was based on {\it HST}/WFPC2 $u_{\rm F336W}$.
Figure 4 gives the comparison of the sensitivity functions 
between the WFPC2 $u_{\rm F336W}$ (dashed line) 
and WFC3 $u_{\rm F336W}$ (solid line) filters on {\it HST}.
The main peaks of the normalized sensitivity functions 
at $\lambda$ = 3000 -- 4000\,{\small \AA} agree well, 
yet the WFPC2 $u_{\rm F336W}$ filter used in Paper II shows 
the red leak at $\lambda$ $\simeq$ 7200\,{\small \AA}. 
The inset is a zoomed-in plot of $\lambda$ = 6500 -- 8000\,{\small \AA} region 
and highlights the red leak of WFPC2 $u_{\rm F336W}$. 
We find that the absence of the red leak of WFC3 $u_{\rm F336W}$ used in this study 
leads to a non-negligible change in $u$-band colors,
compared to those based on WFPC2 $u_{\rm F336W}$.
Tables 2 and 3 give the model data for $u-z$ and $u-g$ CMRs, respectively,
for 10--14 Gyr with fine grid spacing ($\Delta$[Fe/H] = 0.1). 
The $g-z$ data is identical to, and available from, Table 2 of Paper II.

We note that, for Milky Way and NGC 5128 GCs in Figure 3,
the observational data points of $u-z$ and $u-g$ are obtained 
by converting currently available $U-I$ and $U-B$, respectively. 
Figure 5 gives the relationships of $u-z$ versus $U-I$ 
and $u-g$ versus $U-B$ (solid lines), 
which are derived from the model data (red crosses) for synthetic GCs 
with combinations of age (10 -- 15 Gyr of 0.1 Gyr intervals) 
and [Fe/H] ($-$2.5 -- 0.5 dex of 0.1 dex intervals). 
As demonstrated in the figure, $U-I$ and $U-B$ are 
good proxies to $u-z$ and $u-g$, respectively. 
As a result, the simple linear fits (solid lines) suffice 
over the range of ages and metallicities. 
Also note that, in Figure 3, the data of $u-z$ and $u-g$ for M87 GCs
are the WFC3 $u_{\rm F336W}$ colors 
converted from the WFPC2 $u_{\rm F336W}$ colors.
Table 4 summarizes the references to the observed data 
and the conversion relations used in this study.

\subsection{Projection of Metallicities onto Colors}

Figure 6 compares the modeled and observed color distributions for M84 GCs. 
Our working hypothesis is that the CMRs are inflected 
and create bimodal GC color distributions 
if the underlying metallicity spread is simply broad
as expected from hierarchical merging of numerous (proto-)galaxies.
In the simulations, varying ages and mean metallicity produce
systematic changes in morphology of $g-z$, $u-z$, and $u-g$ model histograms. 
With no a priori knowledge on the shape of the underlying MDF, 
we assume a simple MDF structure of Gaussian normal distribution. 
%As we will show in Section 3.3, the Gaussian distributions 
%are in fact quite consistent with the inferred MDFs by inverse-transformation. 
The input MDF and age are interactively adjusted 
until we reach the best match between modeled and observed color histograms 
for $g-z$, $u-z$ and $u-g$, {\it simultaneously}. 
The combination of parameters that matches up all the morphologies 
of $g-z$, $u-z$, and $u-g$ histograms at the same time 
are ($\langle{\rm [Fe/H]}\rangle$, age) = ($-$0.9 dex, 13.0 Gyr) 
with a fixed $\sigma_{\rm [Fe/H]}$ = 0.6 dex.

To quantify the bimodality properties, 
we use the Gaussian Mixture Modeling (GMM) code by \citet{muratov10}.
The results of the GMM analysis for the modeled and observed histograms in Figure 6 
are given in Table 5, 
under the two assumptions for a distribution to be 
homoscedastic (i.e., a mixture of two normal distributions with the same variance)
or heteroscedastic (i.e, with different variances). 
The table gives the mean color ($\mu$), the standard deviation ($\sigma$), 
and the number fractions ($f$) of blue (subscript $b$) and red (subscript $r$) GCs.
The last three columns list the probabilities of preferring a unimodal distribution 
over a bimodal distribution ($p$-value) 
derived based on the likelihood ratio test ($\chi^{2}$), 
on the separation of the means relative to their variances ($DD$) 
and on the kurtosis of a distribution ($kurt$).

Back in Figure 6, the $g-z$ case is shown in the leftmost column.
The column shows how the inflection on a CMR causes color bimodality 
by projecting equidistant metallicity intervals near the quasi-inflection point 
(i.e., the most metallicity-sensitive point) onto wider color intervals. 
In the second row, as an aid to visualizing the simulated color distributions 
we plot $g-z$ of synthetic GCs against their modeled $u$-band absolute mag, $M_u$.
Even with the observational uncertainties fully taken into account, 
the split between two vertical bands of GCs is readily visible. 
This leads to the dip at the midpoint of the color histogram (third row).
In this way, the nonlinear projection results in bimodal color distributions 
even when the underlying distribution in [Fe/H] is unimodal. 
The agreement in morphologies between the simulated (third row) and observed (bottom row) 
$g-z$ distributions is remarkable.
%albeit with slight offsets in color.
A GMM bimodal fitting (Table 5) gives, for the homoscedastic cases, 
[($\mu_b$, $\mu_r$), ($f_b$, $f_r$)] = [(0.98, 1.44), (62\,\%, 38\,\%)] for the simulated histogram
and [(0.96, 1.35), (63\,\%, 37\,\%)] for the observed one.
The heteroscedastic case yields [(0.92, 1.34), (44\,\%, 56\,\%)] for the simulated histogram
and [(0.94, 1.32), (56\,\%, 44\,\%)] for the observed one. 
%[[ THE KS-TEST HERE ]] 
The result suggests that a color distribution of a GC system does not directly reveal its MDF. 
But instead, for a given metallicity spread, the color histograms 
may be primarily determined by the form of the CMR.

The $u-z$ and $u-g$ cases, on the other hand, are shown 
in the middle and rightmost columns of Figure 6, respectively. 
It is clear in the bottom panels that 
the observed $u$-band color distributions are 
significantly different from the observed $g-z$ distribution: 
The prominence of bimodality shown in $g-z$ is 
substantially reduced in $u-z$ and almost diminished in $u-g$. 
The probabilities of preferring a unimodal distribution over a bimodal distribution 
derived based on the separation of the means relative to their variances,
$p$-value($DD$) = 0.00, 0.07, and 0.42 
for $g-z$, $u-z$ and $u-g$ respectively, for the homoscedastic case, 
and 0.12, 0.45, and 0.61 the heteroscedastic case (Table 5). 
The third row presents the modeled histograms. 
The degree of nonlinearity is in order of $g-z$, $u-z$ and $u-g$ (top panels)
and, in turn, the strength of color bimodality is in the same order.
As a consequence, our model predictions (third row) 
based on the nonlinear CMR hypothesis
match up well with the observed distributions (bottom row)
in terms of their overall morphologies.

To be more quantitative, 
we also apply the GMM test to the simulated and observed histograms for $u-z$ and $u-g$ (Table 5). 
For the $u-z$ color (middle column), 
the homoscedastic case gives, 
[($\mu_b$, $\mu_r$), ($f_b$, $f_r$)] = [(2.77, 3.78), (73\,\%, 27\,\%)] for the simulated histogram
and [(2.57, 3.41), (67\,\%, 33\,\%)] for the observed one.
The heteroscedastic case yields 
[(2.55, 3.37), (39\,\%, 61\,\%)] for the simulated histogram
and [(2.44, 3.08), (36\,\%, 64\,\%)] for the observed one. 
For the $u-g$ color (rightmost column), 
the homoscedastic case gives 
[(1.80, 2.45), (86\,\%, 14\,\%)] for the simulated histogram
and [(1.64, 2.12), (79\,\%, 21\,\%)] for the observed one.
The heteroscedastic case yields 
[(1.66, 2.07), (43\,\%, 57\,\%)] for the simulated histogram
and [(1.52, 1.81), (25\,\%, 75\,\%)] for the observed one.
%[[ THE KS-TEST HERE ]] 
We note that the blue peak color of modeled GCs 
are redder than the observation by $\sim$\,0.1 mag in $u-g$.
Our stellar population models show that, for given input parameters, 
the {\it absolute} colors of model GCs can vary up to $\sim$\,0.2 mag in $g-z$, $u-z$ and $u-g$,
depending on the choice of stellar evolutionary tracts and model flux libraries.
We hence put more weight on the {\it relative} color values, 
i.e., the blue and red GC number fraction and the overall morphologies 
of the simulated color histograms.

The metallicity--color nonlinearity provides a good explanation 
for the systematic variation in strength of color bimodality of M84 GC system (Figure 6). 
Nevertheless, it is still important to check whether observational measurement errors
and possible GC-to-GC variations in the physical parameters
play any role in weakening bimodality of $u$-band colors. 
Figure 7 demonstrates how a bimodal [Fe/H] distribution behaves in the color domain
as the color scatter substantially increases to a factor of three. 
In the first row, we assume the conventional linear CMRs
and a bimodal underlying MDF. 
The combination of two Gaussian normal distributions is adapted 
from the [Fe/H]$_{(g-z)}$ histogram in Figure 8 
(dotted histogram in the bottom-left panel). 
%and best reproduce the observed $g-z$ histogram (leftmost panel in the second row).
The second row allows the color scatter 
just as given by the observational uncertainties in color.
The  measurement errors of $g-z$, $u-z$ and $u-g$ 
for each magnitude bin and the entire sample are summarized in Table 6.
For the entire 362 GCs, the median $u-z$ and $u-g$ errors 
are respectively 0.058 and 0.056 mag, 
which are both $\sim$1.4 times larger than 0.041 mag for $g-z$. 
However, the approximate color ranges spanned by $g-z$, $u-z$, and $u-g$ distributions 
are 0.6\,$\sim$\,1.7 mag, 1.9\,$\sim$\,4.3 mag, and 1.0\,$\sim$\,2.7 mag, respectively. 
That is, [$\Delta$($g-z$), $\Delta$($u-z$), $\Delta$($u-g$)] $\simeq$ [1.1, 2.4, 1.7], 
meaning that the baselines of $u-z$ and $u-g$ are 
respectively 2.2 and 1.5 times longer than that of $g-z$. 
As a result, the relative sizes ($\epsilon$) of error bars are calculated 
to be [$\epsilon$($g-z$) : $\epsilon$($u-z$) : $\epsilon$($u-g$)] 
$\simeq$ [1.0\,/\,1.1 : 1.4\,/\,2.4 : 1.4\,/\,1.7] $\simeq$ [0.9 : 0.6 : 0.8]. 
It is important to observe that in a {\it relative} sense, 
the observational uncertainties in three colors 
are quite comparable with one another 
and in fact the errors in $u-z$ and $u-g$ are smaller than that in $g-z$. 
This is evident in the error bars shown in Figure 3 as well.
As a consequence, the modeled $u-z$ and $u-g$ histograms 
(middle and right columns in the second row of Figure 7) 
based on the conventional linear CMRs 
are not consistent with the observations (Figure 6). 
It is, therefore, highly unlikely that bimodal histograms are 
simply blurred by larger observational errors in the $u$-band.

In Figure 7, we also test how extra color uncertainty posed by the GC-to-GC variations 
in terms of the stellar population parameters
(e.g., age, [$\alpha$/Fe], helium abundance, initial mass function, and multiple stellar populations)
exerts effect on the color distributions.
%Obviously, [Fe/H] is not the only parameter determining colors.
Adopting a single relation between metallicities and colors
should underrate the possible spread of color distributions, 
and the intrinsic scatter around the [Fe/H]--color calibrations could be 
a contributor to diluting [Fe/H] distributions in color space.
In order to accommodate such variations, 
the third and bottom rows allow
twice and three times larger scatter than the observational uncertainties.
Although the actual GC-to-GC variations of M84 GC system 
in physical parameters are not known,
the color--color diagrams in Figure 2 give an indication that
extra scatter other than measurement errors 
around the relations is fairly small. 
Our experiment shows that 1.2 times the measurement error 
best reproduce the $g-z$ vs. $u-z$ and the $u-z$ vs. $u-g$ diagrams
and that 1.4 times match the $u-g$ vs. $g-z$ diagram.
So, twice and three times larger scatter
in the third and bottom rows of Figure 7
are well beyond the estimates of GC-to-GC variations of M84 GCs. 
Even with the excessively large scatter in color,
the way a bimodal metallicity distribution manifest itself in color
are not consistent with the observation (Figure 6).
One might argue that the difference in the overall slopes, $\Delta$color/$\Delta$[Fe/H], of the linear fits, 
combined with scatters in color, 
could lead to varying degrees of color bimodality for a bimodal [Fe/H] distribution.
However, the {\it apparent} slopes are turned out to be almost identical (first row) 
because the different $\Delta$color/$\Delta$[Fe/H] slopes 
are fully compensated by  
the the differing color baselines ranged by $g-z$, $u-z$, and $u-g$ distributions, 
i.e., [$\Delta$($g-z$), $\Delta$($u-z$), $\Delta$($u-g$)] $\simeq$ [1.1, 2.4, 1.7].
Therefore, it is highly unlikely that, for M84 GC system,
blurring due to the GC-to-GC scatter
smears out a bimodal MDF in the color domain.

% NOTE THAT color scatter larger for g-z for given amount of variations

%A KolmogorovÐSmirnov test on the mass functions shows a $>$98 per cent probability 
%that all three subpopulations are distinct from one another.

We finish this Section by emphasizing that 
in the context of the nonlinear CMRs, 
the systematic variation in the histogram morphology for different colors 
is readily explained if the shape of the CMRs is subject to colors (Figure 6). 
By contrast, in the conventional view established based on the linear CMRs, 
the color histogram morphology has no reason to vary depending on the colors,
unless the scatter are significantly different color-by-color.
That is, two distinct GC subpopulations should manifest themselves 
more or less in the same way even in different color distributions (Figure 7). 
We therefore conclude for M84 GC system that 
not only it is unnecessary to assume a bimodal [Fe/H], 
but also it is more appropriate to assume a unimodal [Fe/H].

\subsection{De-projection of Colors onto Metallicities}

In an attempt to investigate the possibility 
of using the theoretical CMRs as a tool for recovering the MDFs, 
this Section performs the de-projection of the observed color distributions of M84 GCs. 
The metallicity-to-color projection carried out in Section 3.2 should be reversible, 
but the inverse transformation can be negatively affected by the following two factors. 
First, compared to the metallicity-to-color transformation, 
the inverse transformation is more susceptible to 
the incompleteness of current population synthesis models. 
As mentioned in Section 3.2, the different choice of 
stellar evolutionary tracks and model flux libraries can result in 
up to $\sim$\,0.2 mag variation in $g-z$, $u-z$, and $u-g$ for given input parameters.
Supposing that a CMR is erroneously shifted in the {\it color} direction, 
the metallicity-to-color conversion would still give correct color histogram morphologies, 
yet the inverse conversion would yield incorrect MDFs. 
Second, the color-to-metallicity de-projections can also be hampered 
by the color uncertainty due to the observational measurement errors and the intrinsic GC-to-GC variations. 
Obviously, the uncertainty makes the histograms of colors broader than the intrinsic distributions
and thus leads to the inferred MDFs that are broader than the true MDFs.
This effect, when combined with the steepness of the metal-poor part of the nonlinear CMRs, 
can result in the enhanced frequency of GCs 
at the very metal-poor tails ([Fe/H] $<$\,$-2.5$).
Thus, the exact shape of the metal-poor wings of inferred MDFs 
should be viewed with caution.
Nevertheless, careful comparison of the GC MDFs obtained 
independently from various color histograms 
will shed light on the color--metallicity nonlinearity issue.

Figure 8 shows the inferred MDFs as products 
of de-projection of the observed colors onto metallicities, 
using both nonlinear and linear CMRs. 
%under the two assumptions for the distributions to be homoscedastic or heteroscedastic. 
%The table gives the mean [Fe/H] ($\mu$), standard deviation ($\sigma$), and the number fractions ($f$) 
%of metal-poor (subscript $mp$) and metal-rich (subscript $mr$) GCs, and the $p$-values.} 
The top row is identical to the bottom row of Figure 6, 
showing the observed $g-z$, $u-z$ and $u-g$ distributions. 
The bottom row presents the GC metallicity histograms 
obtained independently from the $g-z$, $u-z$ and $u-g$ colors 
based on the theoretical CMRs shown in the middle row. 
The best-fit age (13 Gyr) for M84 GCs derived in Figure 6 is also used here.
The observed color histograms with different morphologies (top row) 
are all converted to the MDFs (filled histograms in the bottom row) 
that have a strong metal-rich peak with a wing on the metal-poor side. 
Although the inverse conversion inevitably overestimates
the metal-poor tail of inferred MDFs,
the three inferred metallicity histograms are consistent with one another 
in terms of their overall shape and peak metallicities at [Fe/H] $\simeq$ $-0.9$.\footnote{
The result of the GMM analysis for the inferred MDFs in Figure 8 is listed in Table 7.
The metal-rich (subscript {\it mr}) fractions, $f_{mr}$, are 98\,\%, 98\,\%, and 97\,\% 
with metal-rich mean [Fe/H], $\mu_{mr}$ =  $-1.0$, $-1.1$, and $-1.3$ 
for the MDFs derived from $g-z$, $u-z$, and $u-g$, respectively, for the homoscedastic case.
The values for $f_{mr}$ are 86\,\%, 85\,\%, and 82\,\% 
with $\mu_{mr}$ =  $-0.9$, $-1.0$, and $-1.0$, for the heteroscedastic case (Table 7).}
By contrast, the distributions (dotted histograms in the bottom row) 
derived based on the conventional {\it linear} color--metallicity relations 
(dotted lines in the middle row) do not agree with one another.\footnote{ 
The values for $f_{mr}$ are 38\,\%, 33\,\%, and 21\,\% 
with metal-rich mean [Fe/H], $\mu_{mr}$ =  $-0.4$, $-0.5$, and $-0.5$ 
for the MDFs derived from $g-z$, $u-z$, and $u-g$, respectively, for the homoscedastic case.
The values for $f_{mr}$ are 42\,\%, 68\,\%, and 30\,\% 
with $\mu_{mr}$ =  $-0.4$, $-0.9$, and $-0.7$, for the heteroscedastic case (Table 7).}
We emphasize that under the nonlinear-CMR assumption, 
the model invariably obtains very similar morphologies 
and peak positions of the MDFs from different colors, 
strongly indicating that the color--metallicity nonlinearity is real.

Figure 9 presents the same de-projection experiment as in Figure 8 with the identical 
observational data, but here the observed color histograms (top row) are broken into two hypothetical subgroups
following the conventional notion.
The distributions of blue and red GCs are fit with a pair of Gaussian functions. 
A GMM bimodal fitting (Table 5) gives 
the mean colors and the number fractions of the blue and red subpopulations as  
[($\mu_b$, $\mu_r$), ($f_b$, $f_r$)] = [(0.96, 1.35), (63\,\%, 37\,\%)] for $g-z$,
[(2.57, 3.41), (67\,\%, 33\,\%)] for $u-z$, and
[(1.64, 2.12), (79\,\%, 21\,\%)] for $u-g$.
We use the GMM analysis for the same variance case, 
but the use of different variances does not affect our conclusion.
The two Gaussian functions and their sums in the top row are 
converted to MDFs through the nonlinear color-to-metallicity 
conversions in the middle row. In the bottom row, the black, 
blue and red histograms are respectively obtained from the 
corresponding curves in the top row. 
Under the nonlinear-CMR assumption, 
the three color distributions with dissimilar blue-to-red GC ratios
are all transformed into simple, broad MDFs with the similar morphologies. 
The broad MDF of the M84 GC system is in accordance with 
the hierarchical merging paradigm of galaxy formation,
and two hypothetical subgroups are not necessarily required.
Therefore, the test performed in Figure 9 
gives further support for the nonlinear-CMR scenario for the color bimodality.

%=========================================================================

\section{DISCUSSION}

We have presented archival {\it HST}/WFC3 F336W ($u$-band) photometry of the GC system in M84, 
against which our simulated GC color distributions are compared. 
The agreement between the observation and simulations strengthens the view 
that the metallicity--color nonlinearity effect has a key role in producing color bimodality.

This Section discusses the implications 
regarding the nonlinear-CMR scenario for GC color bimodality (Section 4.1) 
and elliptical galaxy formation theories (Section 4.2).
Here we combine the result on M84 with that of M87 presented in Paper II 
to carry out a comparative examination of their GC systems.
M87 and M84 represent galaxies situated in a cluster of galaxies,
and M87, compared to M84, is an example of a more massive galaxy 
closer to the heavily populated inner core of a galaxy cluster. 
Note that the GC samples for M87 (with {\it HST} ACS/WFC and WFPC2) 
and M84 (with {\it HST} ACS/WFC and WFC3) are confined to $R \lesssim R_{\rm eff}$ 
at the Virgo distance ($\simeq$ 16.5 Mpc).

\subsection{Globular Cluster Systems of M87 and M84 as Testbeds of the Metallicity--Color Nonlinearity Scenario} 

There is extensive ongoing debate as to,
between the metallicity difference and the nonlinear projection effect,
which plays the more important role in making GC color bimodality.
To address this issue,
Figure 10 presents the observed and simulated color distributions of 
M87 (Paper II) and M84 (this study) GC systems. The upper six panels 
compare the observed histograms for the $g-z$, $u-z$, and $u-g$ colors 
of M87 (upper row) and M84 (lower row) GCs. Both galaxies show that 
bimodality in the $g-z$ distribution (leftmost panels) is reduced in $u-z$ (middle) and 
further weakened in $u-g$ (rightmost), in a very systematic manner. When one 
compares the two galaxies in the given colors, the red peak of M84 is 
significantly weaker than that of M87 for every color histogram, 
again, in a very systematic way.

The lower six panels show that the orderly behavior of the observed color distributions 
is reproduced well by the simulated histograms
from $g-z$ to $u-z$ to $u-g$ for a given galaxy 
and from M87 to M84 for a given color.
The best-fit parameters for M87 and M84 are summarized in Table 8, 
along with their basic observational information.
As described in Section 3.2, the model needs only two adjusting parameters, 
i.e., the mean [Fe/H] ($\langle$[Fe/H]$\rangle$) with a fixed $\sigma_{\rm [Fe/H]}$ and age ($t$). 
The parameters are selected to be ($\langle$[Fe/H]$\rangle$, $t$) = ($-0.5$ dex, 13.9 Gyr) 
and ($-0.9$ dex, 13.0 Gyr) for the M87 and M84 GC systems, respectively.

The fact that all the simulations in Figure 10 are performed under the assumption 
of the unimodal [Fe/H] spread greatly reduces the conventional demand for the 
two GC subgroups to explain the color bimodality. 
It is therefore suggested that the nonlinear CMRs are truly universal for old GC systems.
Interestingly, although hampered by the possible presence of young GC populations ($\ll$ 10 Gyr), 
there are some GC systems showing bimodal spectroscopic metallicity
(e.g., NGC 3115 (S0 type) in Brodie et al. 2012 and Usher et al. 2012, but see also Paper V). 
In such systems, the inflected CMRs often create stronger bimodality in the color domain.

The results on the GC systems of the two giant galaxies in the Virgo 
galaxy cluster lend confidence to the effectiveness of the 
``$u$-band'' technique to potentially judge whether the true form of 
CMRs are linear or nonlinear. In this regard, the {\it HST}/WFC3 
observations in F336W for nearby large elliptical galaxies are highly 
anticipated. If the nonlinearity of CMRs is found to be favored by 
the future observations, it will change much of the current thought 
on the GC color bimodality as well as the formation of GCs and their 
host galaxies.

\subsection{Globular Cluster Systems of M87 and M84 as Tracers of Formation of their Host Galaxies} 

Clusters can place important constraints on the histories of 
chemical enrichment and star formation of their parent galaxies.
Figure 11 presents the inferred MDFs of 
M87 (Paper II) and M84 (this study) GC systems. 
As described in Section 3.3, the distributions of 
[Fe/H]$_{(g-z)}$, [Fe/H]$_{(u-z)}$, and [Fe/H]$_{(u-g)}$ were
independently derived respectively from the $g-z$, $u-z$, and $u-g$ histograms 
via the nonlinear color-to-[Fe/H] de-projection.
Each [Fe/H] distribution is an average of the three histograms of 
[Fe/H]$_{(g-z)}$, [Fe/H]$_{(u-z)}$ and [Fe/H]$_{(u-g)}$ for each galaxy. 
Despite the clear difference in the observed GC color histograms between M87 and M84 (Figure 10), 
the nonlinear inverse conversions result in the similar MDF shapes, 
which are characterized by a sharp metal-rich peak with a metal-poor wing.

Recently, Paper III demonstrated that 
the strongly-peaked, unimodal MDFs with broad metal-poor tails 
are similar to the MDFs of resolved halo stars in nearby elliptical galaxies, 
e.g., the M87 field-star MDF reported by Bird \etal\ (2010). 
The characteristic form of the MDFs of both GCs and field stars 
may have a profound implication 
because the unimodal, skewed MDFs are 
products of rather continuous chemical evolution. 
The strongly peaked GC MDFs are consistent with 
hierarchical formation theories of giant elliptical galaxies, 
in which an aggregate of a large number of protogalactic gas clouds 
forms stars and GCs on a relatively short timescale. 
Furthermore, the mutual similarity in the MDF shape between M87 and M84 GC systems (Figure 10)
suggests a common process of GC formation and evolution between the two systems.

It is interesting that while the two MDFs have similar shape, 
they have differing peak positions 
at [Fe/H]$_{peak}$ = $-0.5$ and $-0.9$ for M87 and M84, respectively. 
That the M87 GC MDF is more metal-rich than the M84 GC MDF implies that 
the chemical enrichment in M87 is more processed than that in M84. 
Another difference between the two GC systems lies in their inferred ages, 
in the sense that M87 appears older than M84 by $\sim$\,1 Gyr
(see Paper III for a detailed discussion of the systematic age difference among GC systems).
The possible age difference may be reminiscent of the galaxy downsizing paradigm 
whereby stars of dimmer galaxies, on average, formed later than those of brighter galaxies. 
This notion seems to hold for GC systems as well.
That is, GC systems in fainter galaxies are on average younger, 
which was referred to as ``downsizing of GC systems'' in Paper III.

Combining the MDF and age arguments above together,
we paint a picture in which M87 as a more massive galaxy in a denser environment 
had earlier GC formation and more efficient chemical enrichment than M84.
We therefore propose a scenario that the formation and metal enrichment of GCs 
have started earlier and proceeded more efficiently in more massive galaxies 
in denser environments (see Paper III for more details). 
In this view, the most metal-poor GCs in most massive galaxies in galaxy clusters 
may serve as the first generation of GCs in the universe.

%=========================================================================
%=========================================================================
\acknowledgments 
We would like to thank the anonymous referee for his/her helpful comments.
S.-J.Y. acknowledges support 
from International Exchange Program for University Researchers (NRF-2011-013-C00031) 
and from Mid-career Research Program (No. 2012R1A2A2A01043870) 
through the National Research Foundation (NRF) of Korea grant funded 
by the Ministry of Education, Science and Technology (MEST), 
and support by the NRF of Korea to the Center for Galaxy Evolution Research (No. 2010-0027910)
and by the Korea Astronomy and Space Science Institute Research Fund 2012 and 2013. 
This work is partially supported by the KASI--Yonsei Joint Research Program (2012--2013) 
for the Frontiers of Astronomy and Space Science funded by the Korea Astronomy and Space Science Institute
and by the DRC program of Korea Research Council of Fundamental Science and Technology (FY 2012).

%=========================================================================
%=========================================================================

\clearpage
%=========================================================================
%% FIGURE 1
%=========================================================================
\begin{figure}
\begin{center}
\includegraphics[width=16cm]{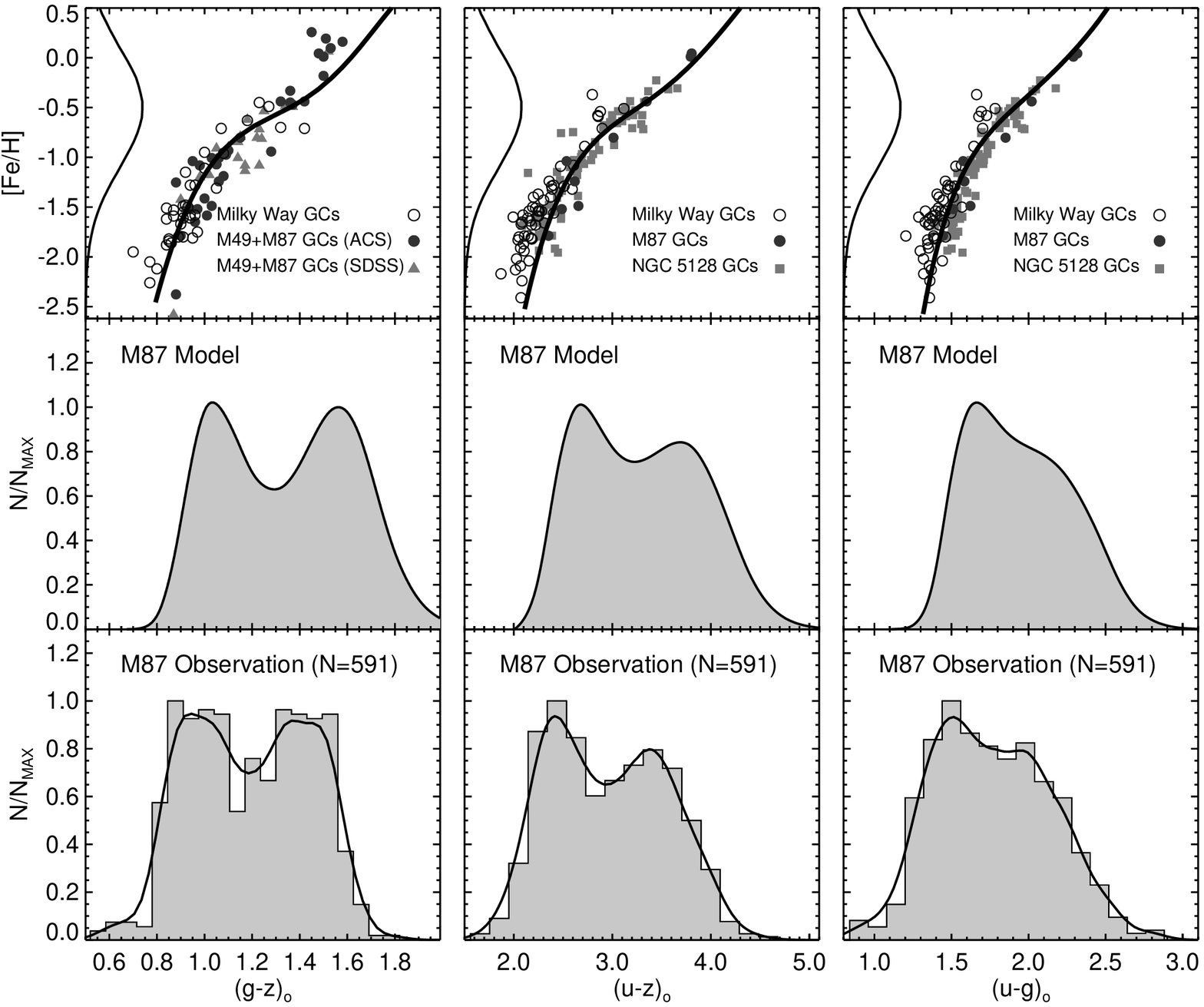}
\caption{\small Mutiband ($u$, $g$, and $z$) photometry of GCs in M87 
and Monte Carlo simulations for their color distributions.
This figure is similar to Figure 2 of Paper II but without color--magnitude diagram panels,
and demonstrates how the metallicity--color nonlinearity works. 
Top row: the observed color--[Fe/H] relations
for Galactic GCs with E($B-V$) $< 0.3$ (open circles),
M49 and M87 GCs with ACS Virgo Cluster Survey photometry (black circles),
M49 and M87 GCs with SDSS photometry (gray triangles),
and NGC 5128 GCs (gray squares). 
Solid curves are for the 5th-order polynomial fit to our model data for the GC age of 13.9 Gyr, 
for which all the morphologies of $g-z$, $u-z$, and $u-g$ color histograms are reproduced {\it simultaneously}. 
The metallicity spread of 10$^6$ model GCs is shown along the y-axis, 
for which a Gaussian normal distribution is assumed 
($\langle$[Fe/H]$\rangle$ = $-0.5$ dex and $\sigma$([Fe/H]) = 0.6).
Middle row: the left, middle, and right columns represent the color distributions 
of 10$^6$ modeled GCs for the $g-z$, $u-z$, and $u-g$ colors, respectively. 
Observational uncertainties as a function of luminosity are taken into account in the simulations.
Bottom row: the observed color histograms for the M87 GC system.
The 591 GCs are used that have reliable measurements in all three bands.
Solid lines are smoothed histograms with Gaussian kernels of 
$\sigma$($g-z$) = 0.05, $\sigma$($u-z$) = 0.15, and  $\sigma$($u-g$) = 0.10.
\label{fig1}}
\end{center}
\end{figure}

%=========================================================================
%% FIGURE 2
%=========================================================================
\begin{figure*}
\begin{center}
\includegraphics[width=16.5cm]{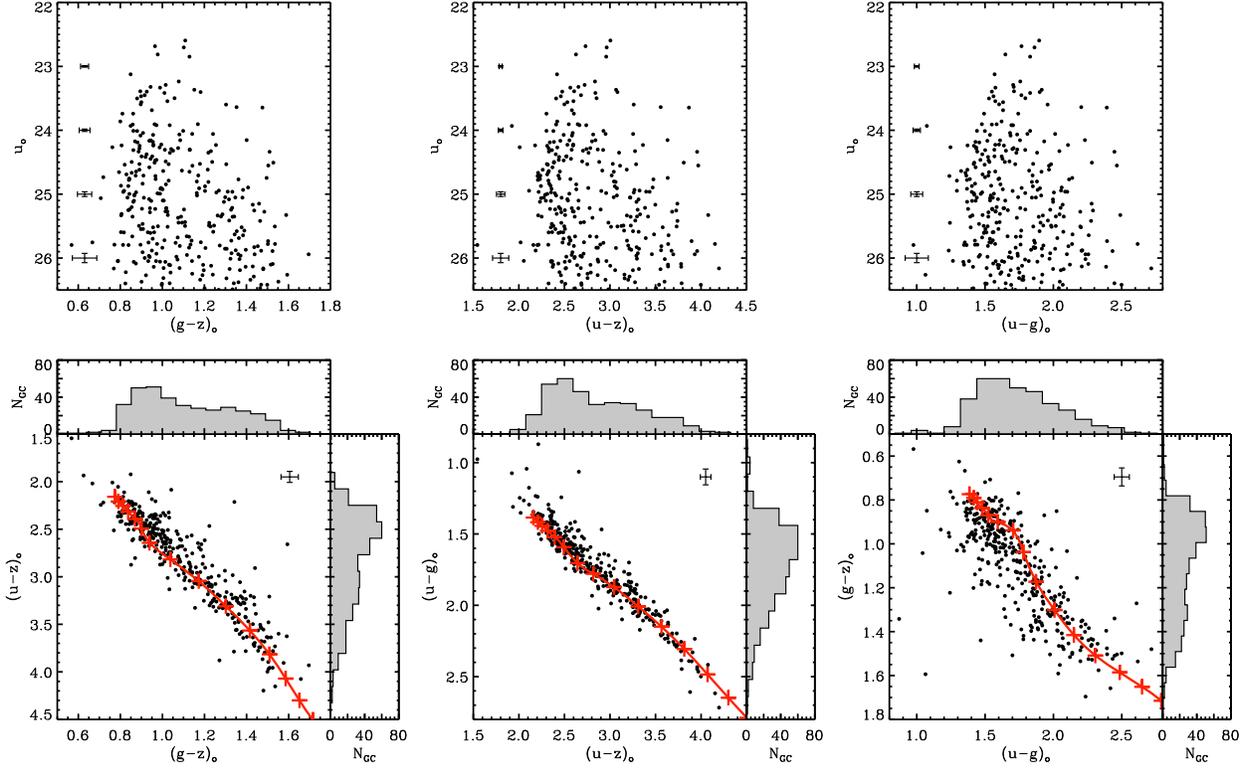}
\caption{The color--magnitude and color--color diagrams for the observed M84 GCs.
This figure is similar to Figure 3 of Paper II, but for M84 GC system (Table 1).
Upper row: the observed color--magnitude diagrams. 
The left, middle, and right panels show the $g-z$, $u-z$, and $u-g$ diagrams, respectively.
The $g$ (ABMAG) and $z$ (ABMAG) mags for GC candidates
are obtained from the ACSVCS GC catalog (Jord\'{a}n \etal\ 2009).
The {\it HST}/WFC3 archival images are used to measure
$u$-band mags (F336W, ABMAG) for the ACSVCS GC catalog. 
A color cut ($g-z$ $>$ 0.5 and $u-g$ $>$ 0.8) was employed 
to filter out contaminating sources such as star-forming background galaxies.
Black dots in each panel are the selected 362 GCs ($\sigma_u$ $<$ 0.2 mag)
that have $u$, $g$, and $z$ measurements in common.
Lower row: the color--color diagrams and the projected color distributions. 
Red solid lines represent our model prediction for 13.0-Gyr GCs (Tables 2 and 3)
from [Fe/H] = $-2.5$ (top left cross) to +0.5 (bottom right).
The red crosses on each model locus mark the uniform [Fe/H] intervals ($\Delta$[Fe/H] = 0.2 dex). 
\label{fig2}}
\end{center}
\end{figure*}

%=========================================================================
%% FIGURE 3
%=========================================================================
\begin{figure*}
\begin{center}
\includegraphics[width=16cm]{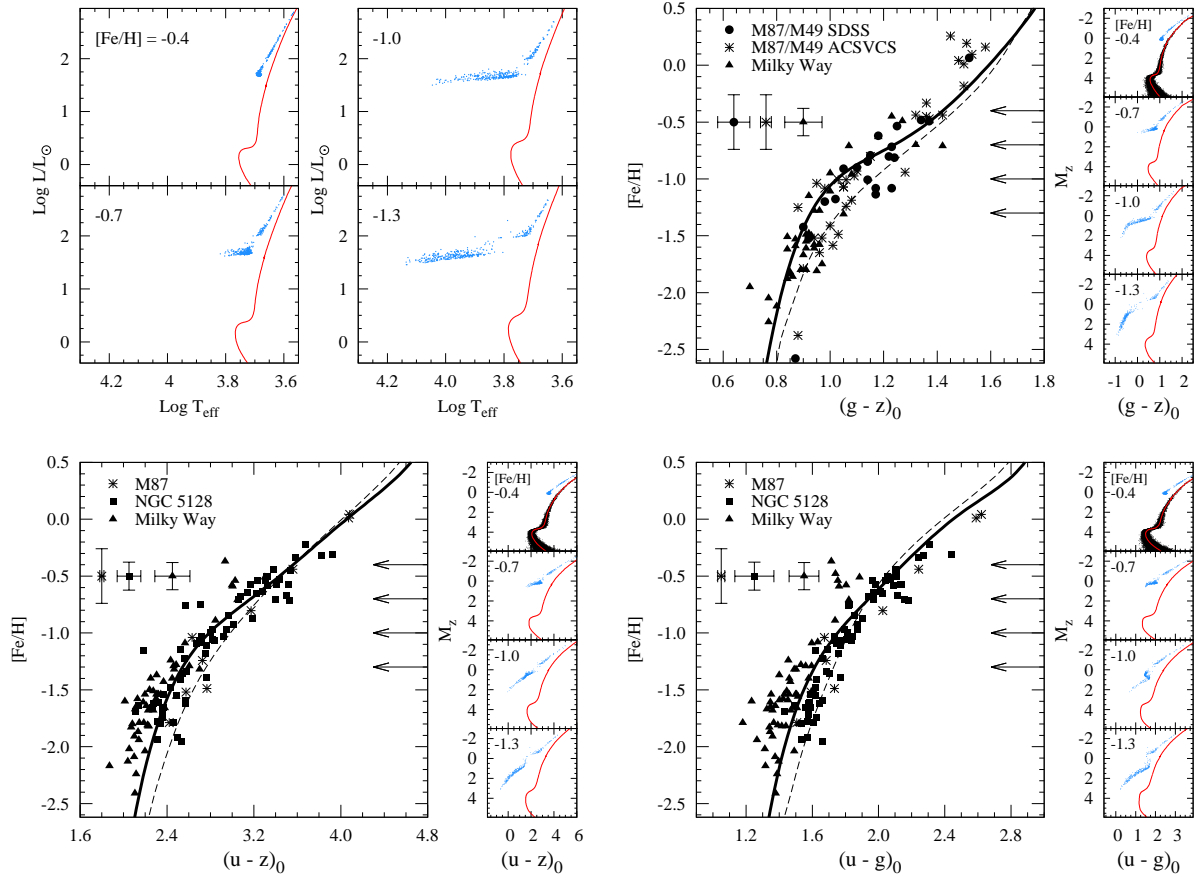}
\caption{The empirical and theoretical CMRs.
This figure is similar to Figure 1 of Paper II, 
but the $u$-band here is {\it HST}/WFC3 $F336W$ (see Figure 4) and the 13-Gyr model is used.  
Upper left quadrant: 
synthetic Log {$T_{\rm eff}$}--Log $L/L_{\sun}$ diagrams 
of individual stars for 13-Gyr model GCs with various [Fe/H]'s.
Red loci are main sequence and giant branch 
from Yonsei-Yale (Y$^2$) isochrones (Kim \etal\ 2002), 
whereas blue dots represent HB stars generated 
based on the Y$^2$ HB tracts (S. Han \etal\ 2012, in preparation).
Upper right quadrant: the ($g-z$)--[Fe/H] CMR.
The information on the data are summarized in Table 4. 
The thick solid line is for the 5th-order polynomial fit to our model data for 13-Gyr GCs,
and the dashed line is for the model without inclusion of the HB prescription. 
%The model $g-z$ data is identical to and available from Table 2 of Paper II. 
Arrows mark the four values of [Fe/H], for which the synthetic color-magnitude 
diagrams are shown in the small panels on the right. 
Synthetic color--magnitude diagrams are generated from the synthetic Log {$T_{\rm eff}$}--Log $L/L_{\sun}$ diagrams 
(upper left quadrant) using the BaSeL flux library (Westera \etal\ 2002).
The top panel shows individual stars (black and blue dots) with error simulations, 
whereas the rest panels show only the corresponding isochrones (red loci) and HB stars (blue dots).
Lower left quadrant: 
the same as the upper right quadrant, but for the $u-z$ color.
%The $u-z$ colors of the GCs in the Milky Way and NGC 5128 
%were converted from their $U-I$ colors (see Table 4 and Figure 5). 
Lower right quadrant: 
the same as the lower left quadrant, but for the $u-g$ color.
\label{fig3}}
\end{center}
\end{figure*}

%=========================================================================
%% FIGURE 4
%=========================================================================
\begin{figure}
\begin{center}
\includegraphics[width=13cm]{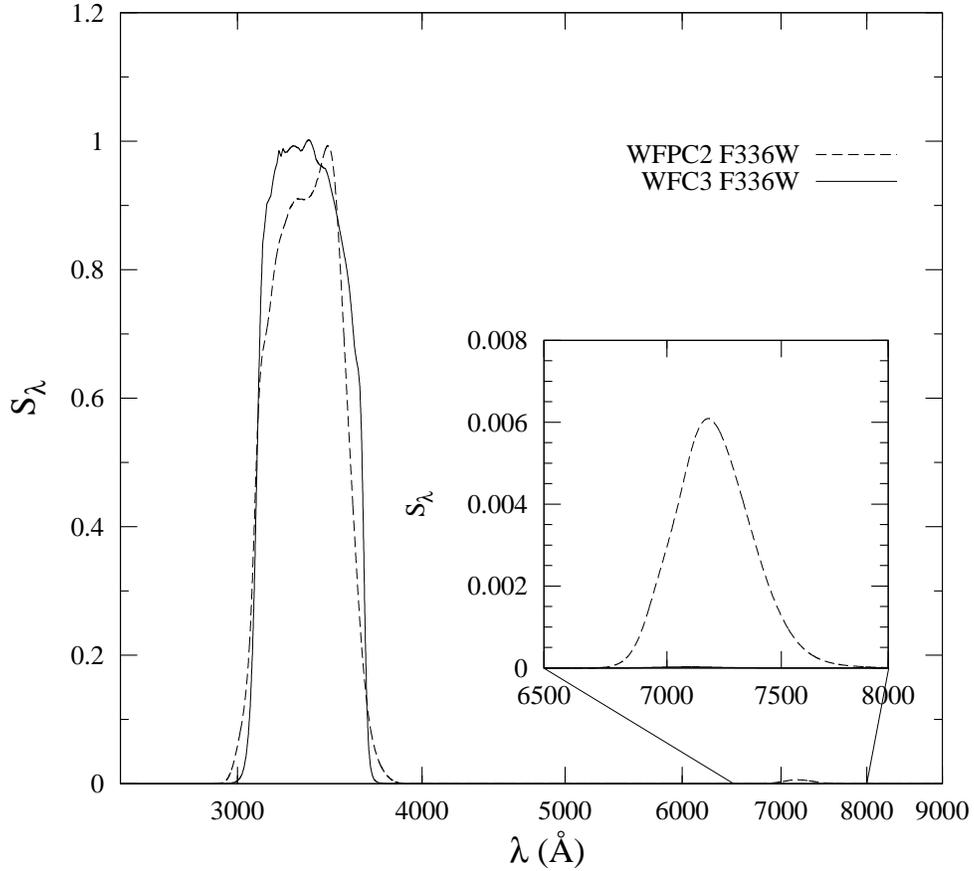}
\caption{Comparison of normalized sensitivity between the WFPC2 $u_{\rm F336W}$ filter (dashed line, Paper II) 
and WFC3 $u_{\rm F336W}$ filter (solid line, this study) on {\it HST}.
The inset is a zoomed-in plot of the red leak of the WFPC2 $u_{\rm F336W}$ filter
at $\lambda$ = 6700 -- 7800\,{\small \AA}. 
The absence of the red leak of WFC3 $u_{\rm F336W}$ leads to a non-negligible change in $u$-band colors 
such as $u-z$ and $u-g$, compared to WFPC2 $u_{\rm F336W}$.
\label{fig4}}
\end{center}
\end{figure}

%=========================================================================
%% FIGURE 5
%=========================================================================
\begin{figure}
\begin{center}
\includegraphics[width=16cm]{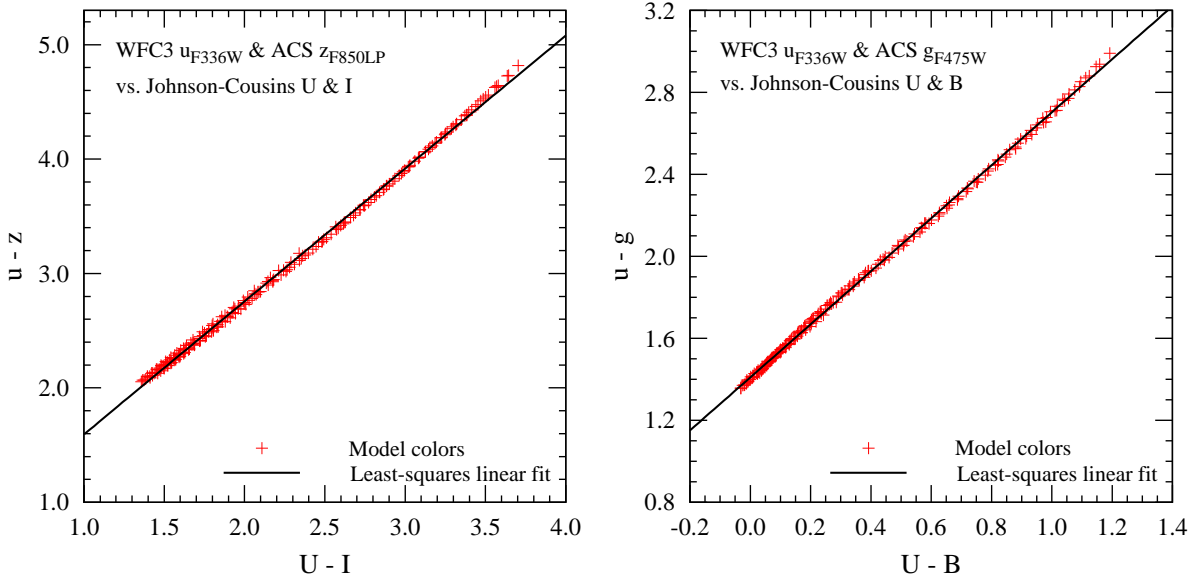}
\caption{Relationships of $u-z$ versus $U-I$ ({\it left panel}) and $u-g$ versus $U-B$ ({\it right}). 
The $u$-band is {\it HST}/WFC3 $u_{\rm F336W}$, for which M84 {\it HST} images are available.
Red crosses represent the model data for synthetic GCs 
with combinations of age (10 -- 15 Gyr of 0.1 Gyr intervals) and [Fe/H] ($-2.5$ -- 0.5 dex of 0.1 dex intervals). 
Least-squares fits (solid lines) to the model data give the two linear equations (see Table 4):
($u-z$) = 1.160 ($U-I$) + 0.434 and ($u-g$) = 1.296 ($U-B$) + 1.412.
\label{fig5}}
\end{center}
\end{figure}

%=========================================================================
%% FIGURE 6
%=========================================================================
\begin{figure*}
\begin{center}
\includegraphics[width=14.5cm]{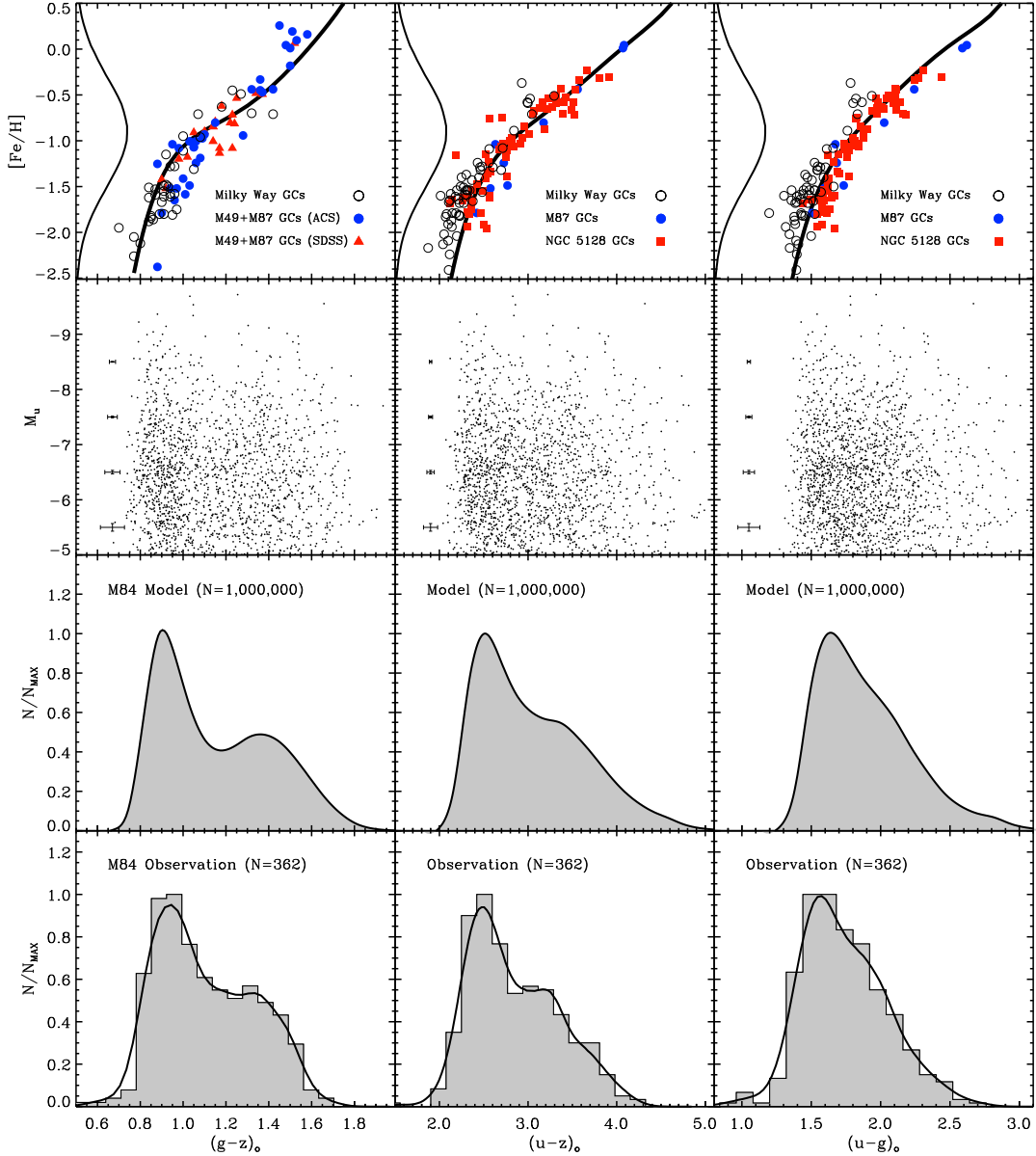}
\caption{\scriptsize Mutiband ($u$, $g$, and $z$) observation of GCs in M84 
and Monte Carlo simulations for their color distributions. 
Top row: the same as the CMRs in Figure 2. 
The metallicity spread of 10$^6$ model GCs is shown along the y-axis, 
for which a Gaussian normal distribution is assumed.  
The best-fit metallicity and age to reproduce the morphologies 
of $g-z$, $u-z$, and $u-g$ color histograms {\it simultaneously} is 
$\langle$[Fe/H]$\rangle$ = $-0.9$ with $\sigma$([Fe/H]) = 0.6 and 13.0 Gyr, respectively.
Second row: the left, middle, and right columns 
represent the color--magnitude diagrams of 2000 randomly selected model GCs
for the $g-z$, $u-z$, and $u-g$ colors, respectively.
The colors are transformed from [Fe/H]'s by using the theoretical relation shown in the top row.
For the integrated $u$-band absolute mag, $M_u$, 
a Gaussian luminosity distribution 
($\langle$$u$$\rangle$ = 25.2, $\sigma$($u$) = 1.15, and distance modulus ($u-M_u$) = 31.33) 
is assumed according to the observation. 
Observational uncertainties as a function of $M_u$ shown by error bars 
are obtained from the actual observations (Table 6) and taken into account in the simulations.
Third row: the left, middle, and right columns 
represent the color distributions of 10$^6$ modeled GCs for the $g-z$, $u-z$, and $u-g$ 
colors, respectively. 
Bottom row: the same as the third row, but the {\it observed} color histograms for the M84 GC system.
The 362 GCs were used that have $u$, $g$, and $z$ measurements in common. 
Solid lines are smoothed histograms with Gaussian kernels of 
$\sigma$($g-z$) = 0.06, $\sigma$($u-z$) = 0.14, and  $\sigma$($u-g$) = 0.10, respectively.
\label{fig6}}
\end{center}
\end{figure*}

%=========================================================================
%% FIGURE 7
%=========================================================================
\begin{figure*}
\begin{center}
\includegraphics[width=13cm]{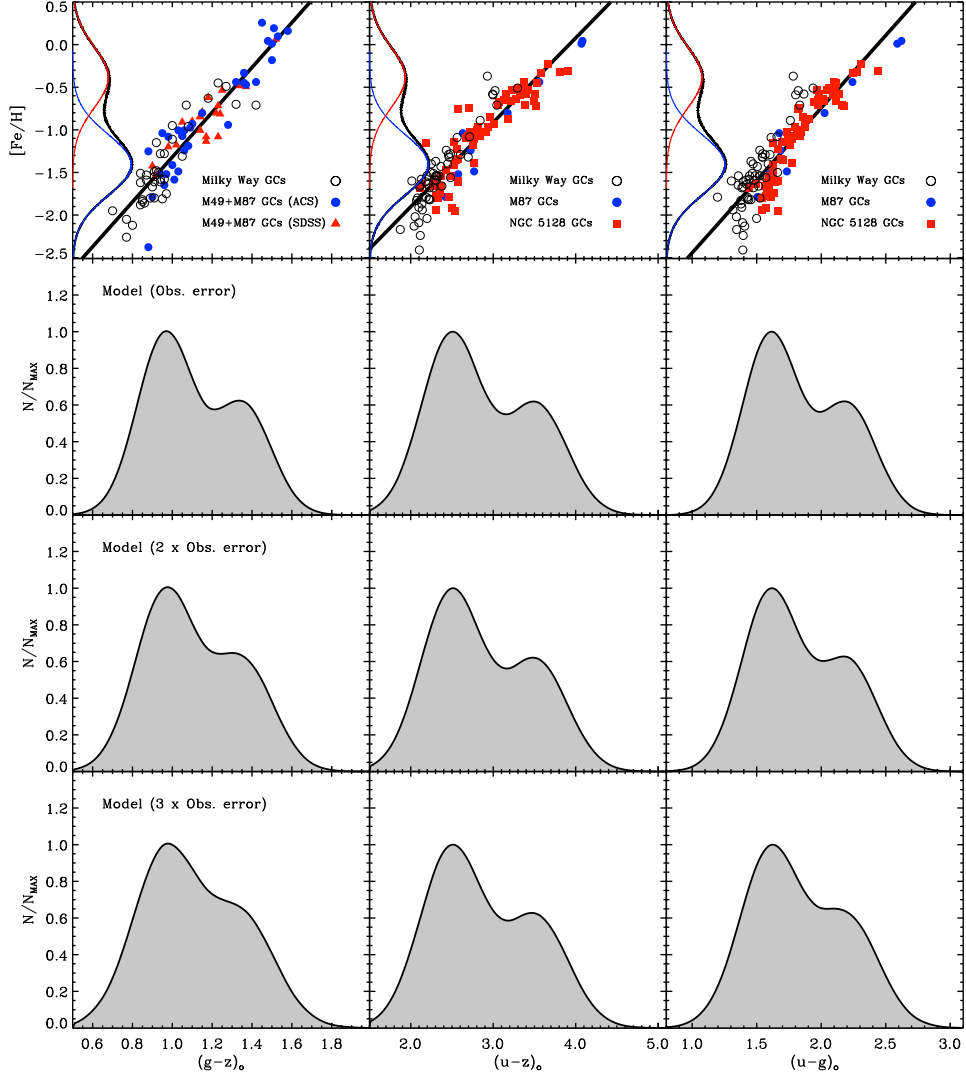}
\caption{\small Similar to Figure 6, but the conventional linear CMRs
and a combination of two Gaussian distributions are assumed. 
Top row: the solid lines represent the least-squares fits to the observational data points. 
The linear equations are [Fe/H] = 2.630\,($g-z$) $-$ 3.950 for $g-z$, 
[Fe/H] = 0.990\,($u-z$) $-$ 3.882 for $u-z$, and 
[Fe/H] = 1.702\,($u-g$) $-$ 4.146 for $u-g$.
The metallicity distribution of 10$^6$ model GCs is shown along the y-axis, 
for which two Gaussian normal distributions 
are adapted from Figure 8 (dotted line in the bottom-left panel).
The mean [Fe/H] and its dispersion, and the number fraction,
($\langle$[Fe/H]$\rangle$, $\sigma$([Fe/H]), $f$),
are ($-$1.41 dex, 0.36 dex, 62\,\%) for the metal-poor group (blue lines)
and ($-$0.39 dex, 0.36 dex, 38\,\%) for the metal-rich group (red lines).
The sum of the two distributions is given by the black line. 
Second row: Monte Carlo simulations for the color distributions. 
The left, middle, and right columns represent the color distributions of 10$^6$ modeled GCs 
for the $g-z$, $u-z$, and $u-g$ colors, respectively. 
The color scatter just as given by the observational uncertainties in color (Table 6) is used.
Third and bottom rows: the same as the second row, but
twice and three times larger scatters than the observational uncertainties are used.
\label{fig7}}
\end{center}
\end{figure*}

%=========================================================================
%% FIGURE 8
%=========================================================================
\begin{figure*}
\begin{center}
\includegraphics[width=15cm]{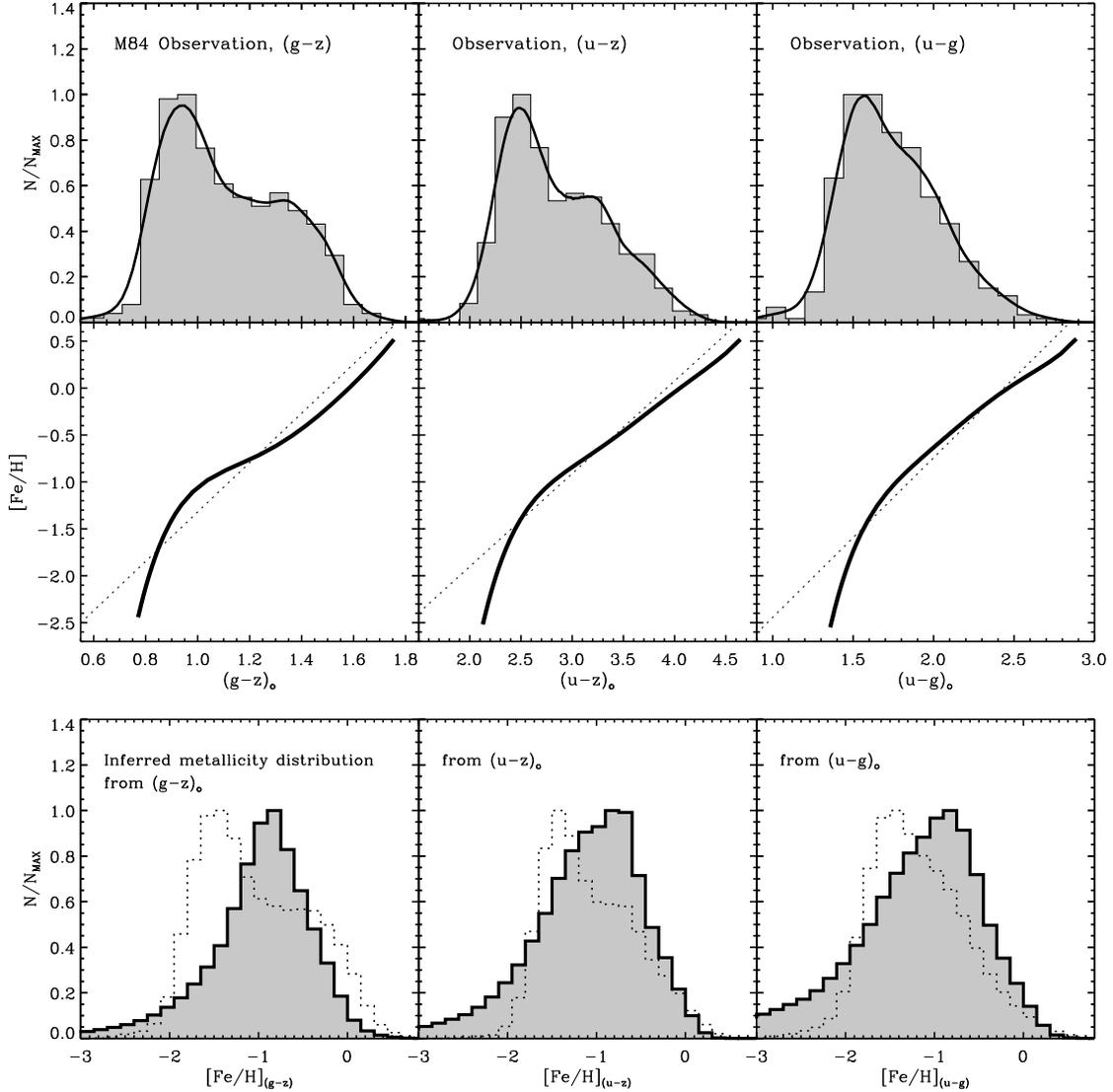}
\caption{The observed $g-z$, $u-z$, and $u-g$ color distributions and their inferred MDFs 
for the GC system of M84. 
Top row: the same as the bottom row of Figure 6. 
Middle row: the same as the top row of Figure 6, 
but the dotted lines represent the least-squares fits to the observational data points 
that are not shown here for clarity (see Figure 7). 
Bottom row: the left, middle, and right panels show the [Fe/H]$_{(g-z)}$, [Fe/H]$_{(u-z)}$, and [Fe/H]$_{(u-g)}$ distributions 
obtained independently from the $g-z$, $u-z$, and $u-g$ color distributions, respectively.
The MDFs are derived from the smoothed histograms in the top row 
via the nonlinear color-to-metallicity transformations (gray histograms with solid lines)
and via the linear transformation (open histograms with dotted lines). 
\label{fig8}}
\end{center}
\end{figure*}

%=========================================================================
%% FIGURE 9
%=========================================================================
\begin{figure*}
\begin{center}
\includegraphics[width=15cm]{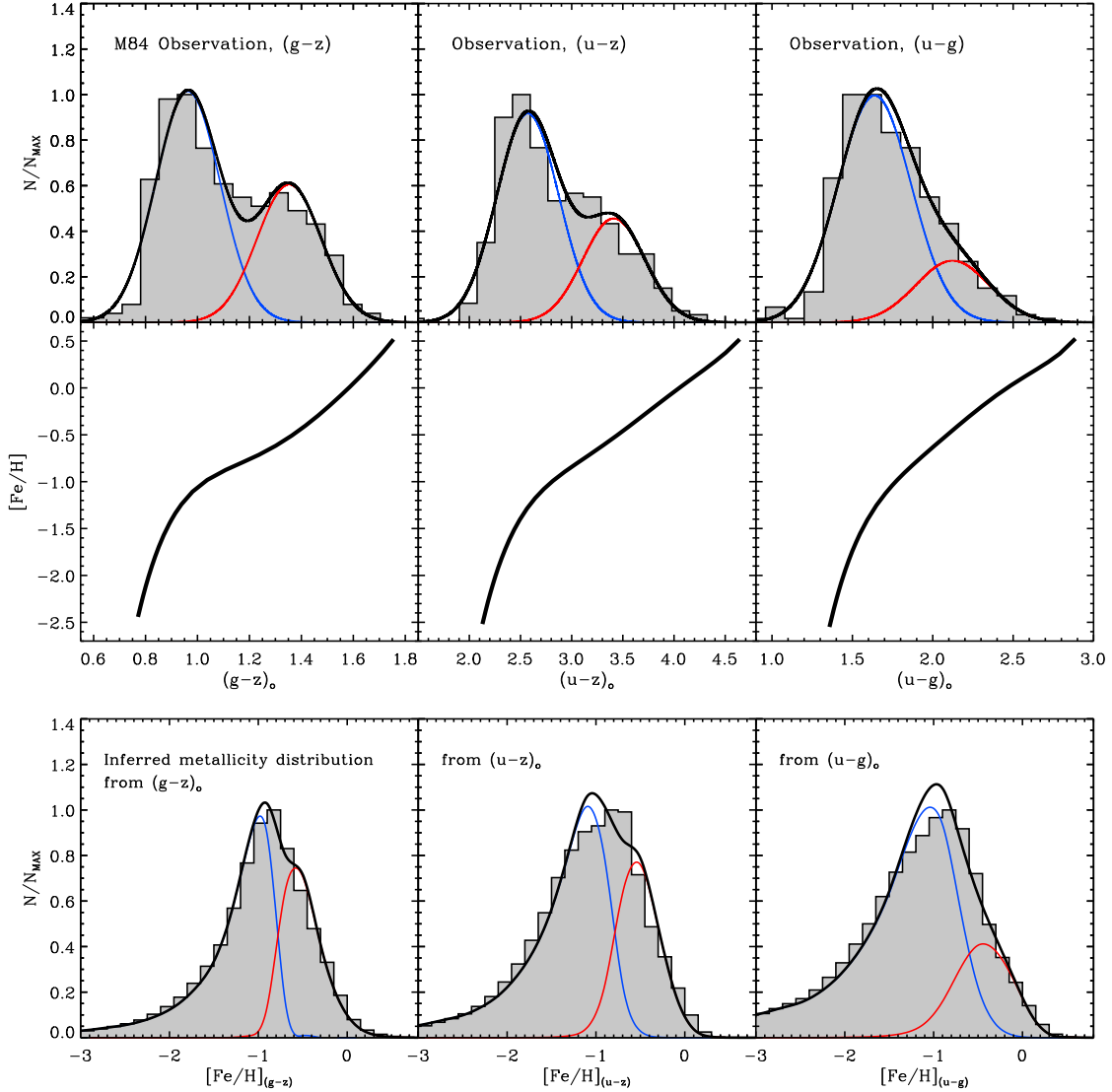}
\caption{The observed $g-z$, $u-z$, and $u-g$ color distributions and their inferred MDFs 
for the GC system of M84.
This figure is the same as Figure 8, but the observed histograms are 
expressed by a sum of two (i.e., blue and red) Gaussian distributions.
Top row: the result of the GMM analysis.
The homoscedastic (same variance) case are used for the GMM test.
The blue, red, and black lines represent the blue, red, and total GCs,
with the peak colors and number fractions of blue and red GCs being
[(0.96, 1.35), (63\,\%, 37\,\%)] for $g-z$,
[(2.56, 3.41), (66\,\%, 34\,\%)] for $u-z$, and
[(1.63, 2.12), (78\,\%, 22\,\%)] for $u-g$, respectively.
Middle row: the same as the middle row of Figure 8. 
Bottom row: the same as the bottom row of Figure 8, 
but the black, blue, and red lines are respectively obtained from the corresponding curves in the top row
via the nonlinear color-to-metallicity transformations in the middle row. 
\label{fig9}}
\end{center}
\end{figure*}

%=========================================================================
%% FIGURE 10
%=========================================================================
\begin{figure*}
\begin{center}
\includegraphics[width=14cm]{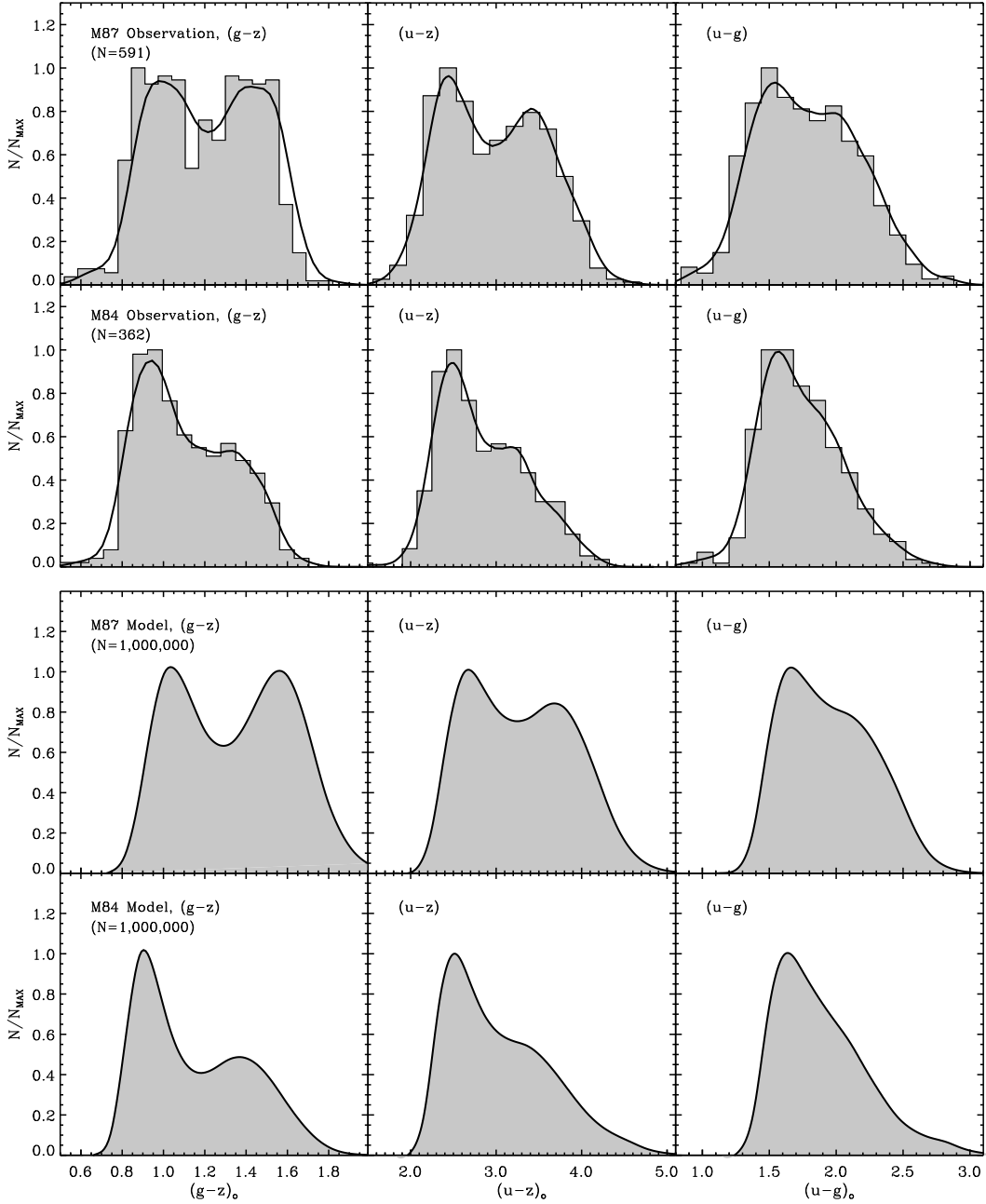}
\caption{\small Comparison between the observed and simulated color distributions 
of M87 (Paper II) and M84 (this study) GC systems. 
The histograms are identical to the ones already shown in Figures 1 (for M87) and 6 (for M84).  
Upper six panels: the observed $g-z$, $u-z$, and $u-g$ color distributions for M87 and M84 GCs.
Note that $u$ corresponds {\it HST}/WFPC2 $F336W$ for M87 and {\it HST}/WFC3 $F336W$ for M84.
Lower six panels: the simulated $g-z$, $u-z$, and $u-g$ color distributions for M87 and M84 GCs. 
The best-fit parameters for M87 and M84 are listed in Table 8.
\label{fig10}}
\end{center}
\end{figure*}

%=========================================================================
%% FIGURE 11 
%=========================================================================
\begin{figure}
\begin{center}
\includegraphics[width=10cm]{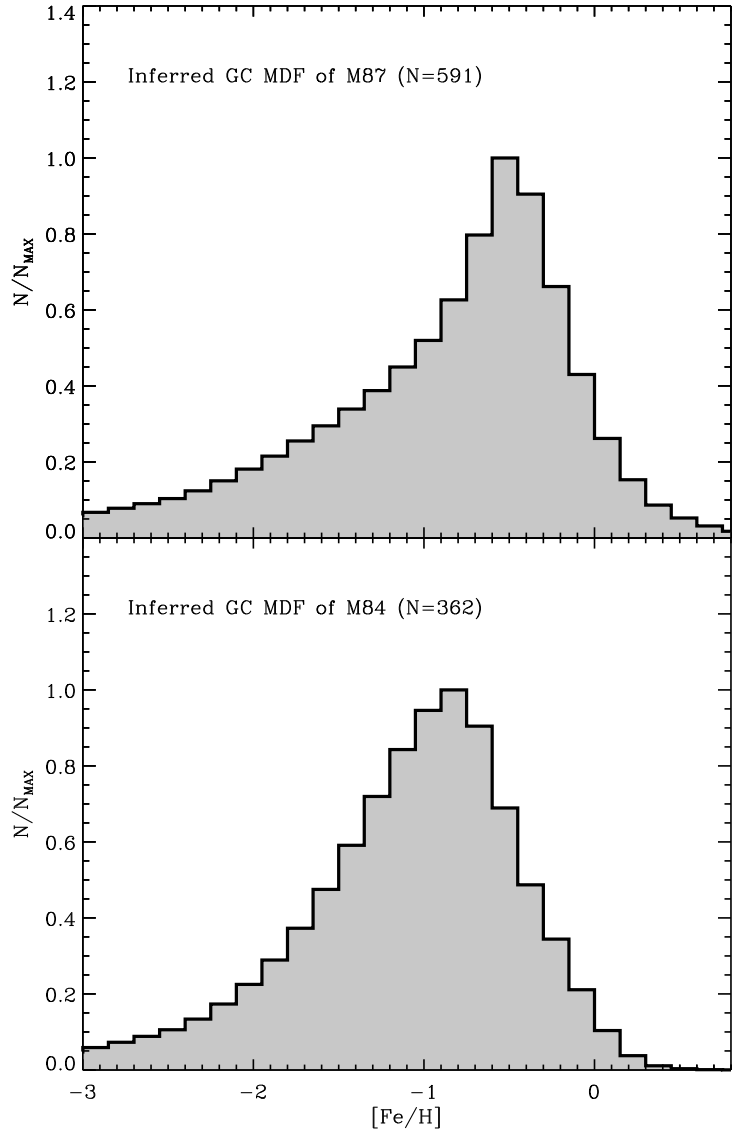}
\caption{Comparison between inferred GC MDFs of M87 (Paper II) and M84 (this study). 
Each MDF is an average of the [Fe/H]$_{(g-z)}$, [Fe/H]$_{(u-z)}$, and [Fe/H]$_{(u-g)}$ histograms (Figure 8)
that are derived respectively from the $g-z$, $u-z$, and $u-g$ color distributions
via the nonlinear color-to-[Fe/H] conversions. 
\label{fig11}}
\end{center}
\end{figure}

%=========================================================================
%= Table 1 ========================================================================
%=========================================================================
\clearpage
\begin{table*}
\begin{center}
\caption{The $u$-, $g$-, and $z$-band Mags and their Observational Errors for the M84 GC Candidates \label{tbl-1}}
\vspace{0.5cm}
\normalsize
\begin{tabular}{c c c c c c c c c}
\hline
\hline
GC ID & RA (J2000) & DEC (J2000) & $u_{0}$ & $u$ Error & $g_{0}$\tablenotemark{a} & $g$ Error\tablenotemark{a} & $z_{0}\tablenotemark{a} $ & $z$ Error\tablenotemark{a}  \\

 & & & \multicolumn{2}{c}{{\scriptsize(WFC3 $F336W$)}} & \multicolumn{2}{c}{{\scriptsize(ACS/WFC $F475W$)}} & \multicolumn{2}{c}{{\scriptsize(ACS/WFC $F850LP$)}}\\
\hline
  1 & 186.2654720 & 12.8857011 & 23.905 & 0.047 & 22.428 & 0.032 & 21.259 & 0.037 \\
  2 & 186.2642216 & 12.8872940 & 23.416 & 0.030 & 21.586 & 0.023 & 20.576 & 0.021 \\
  3 & 186.2640406 & 12.8867801 & 23.932 & 0.043 & 22.858 & 0.014 & 22.009 & 0.071 \\
  4 & 186.2666970 & 12.8860645 & 25.679 & 0.280 & 23.420 & 0.064 & 21.750 & 0.033 \\
  5 & 186.2637346 & 12.8868181 & 25.054 & 0.119 & 23.211 & 0.030 & 21.830 & 0.031 \\
  6 & 186.2644810 & 12.8854047 & 26.143 & 0.205 & 24.772 & 0.125 & 22.918 & 0.107 \\
  7 & 186.2635033 & 12.8870891 & 24.908 & 0.067 & 23.006 & 0.027 & 21.831 & 0.047 \\
  8 & 186.2670468 & 12.8884144 & 24.947 & 0.057 & 23.653 & 0.051 & 22.835 & 0.051 \\
  9 & 186.2675672 & 12.8860944 & 24.496 & 0.041 & 22.851 & 0.035 & 21.614 & 0.041 \\
 10 & 186.2660015 & 12.8845349 & 24.778 & 0.043 & 23.410 & 0.029 & 22.569 & 0.099 \\
\nodata & \nodata & \nodata & \nodata & \nodata & \nodata & \nodata & \nodata & \nodata \\
\nodata & \nodata & \nodata & \nodata & \nodata & \nodata & \nodata & \nodata & \nodata \\
\nodata & \nodata & \nodata & \nodata & \nodata & \nodata & \nodata & \nodata & \nodata \\
\hline
\end{tabular}
\tablecomments{\tablenotemark{a} The $g$- and $z$-band data are obtained from \citet{jordan09}.\\
(This table is available in its entirety in a machine-readable form in the online journal. 
A portion is shown here for guidance regarding its form and content.)}
\end{center}
\end{table*}

%=========================================================================
%= Table 2 ========================================================================
%=========================================================================
\begin{rotate}
\begin{table*}
\tiny
\begin{center}
\caption{\footnotesize Theoretical $u-z$ for Synthetic GCs with Various Ages ($t$) 
Based on the YEPS Model without ($w/o$) and with ($w$) the Inclusion of the HB Prescription 
(The 5th-order Polynomial Fits to the Model Data).\label{tbl-2}}
\begin{tabular}{rcccccccccccccccccc}
\hline
\hline
[Fe/H] & \multicolumn{18}{c}{$u-z$ ($u$ = WFC3/$F336W$, $z$ = ACS/WFC/$F850LP$) } \\
\hline
{} &
\multicolumn{2}{c}{t = 10.0 (Gyr)} &
\multicolumn{2}{c}{10.5} &
\multicolumn{2}{c}{11.0} &
\multicolumn{2}{c}{11.5} &
\multicolumn{2}{c}{12.0} &
\multicolumn{2}{c}{12.5} &
\multicolumn{2}{c}{13.0} &
\multicolumn{2}{c}{13.5} &
\multicolumn{2}{c}{14.0} \\
\hline
& $w/o$ & $w$ & $w/o$ & $w$ & $w/o$ & $w$ & $w/o$ & $w$ & $w/o$ & $w$ & $w/o$ & $w$ & $w/o$ & $w $& $w/o$ & $w$ & $w/o$ & $w$ \\
\hline
--2.5 & 2.171 & 2.078 & 2.189 & 2.078 & 2.208 & 2.078 & 2.224 & 2.086 & 2.240 & 2.093 & 2.254 & 2.112 & 2.268 & 2.132 & 2.280 & 2.155 & 2.293 & 2.178 \\
--2.4 & 2.194 & 2.104 & 2.213 & 2.104 & 2.232 & 2.104 & 2.249 & 2.110 & 2.266 & 2.116 & 2.281 & 2.135 & 2.296 & 2.154 & 2.310 & 2.178 & 2.324 & 2.202 \\
--2.3 & 2.220 & 2.132 & 2.239 & 2.133 & 2.259 & 2.133 & 2.276 & 2.137 & 2.294 & 2.140 & 2.310 & 2.159 & 2.326 & 2.178 & 2.341 & 2.203 & 2.356 & 2.228 \\
--2.2 & 2.247 & 2.163 & 2.267 & 2.163 & 2.287 & 2.163 & 2.305 & 2.165 & 2.323 & 2.166 & 2.340 & 2.185 & 2.357 & 2.203 & 2.373 & 2.229 & 2.390 & 2.255 \\
--2.1 & 2.275 & 2.195 & 2.296 & 2.195 & 2.317 & 2.196 & 2.336 & 2.195 & 2.355 & 2.194 & 2.373 & 2.212 & 2.391 & 2.229 & 2.408 & 2.256 & 2.426 & 2.284 \\
--2.0 & 2.307 & 2.230 & 2.328 & 2.230 & 2.349 & 2.231 & 2.369 & 2.228 & 2.389 & 2.225 & 2.408 & 2.241 & 2.427 & 2.257 & 2.445 & 2.286 & 2.463 & 2.314 \\
--1.9 & 2.341 & 2.268 & 2.363 & 2.268 & 2.385 & 2.269 & 2.405 & 2.263 & 2.426 & 2.258 & 2.446 & 2.274 & 2.465 & 2.288 & 2.485 & 2.318 & 2.504 & 2.346 \\
--1.8 & 2.377 & 2.309 & 2.400 & 2.310 & 2.423 & 2.310 & 2.445 & 2.302 & 2.466 & 2.295 & 2.486 & 2.309 & 2.507 & 2.322 & 2.527 & 2.351 & 2.546 & 2.380 \\
--1.7 & 2.417 & 2.355 & 2.441 & 2.355 & 2.465 & 2.356 & 2.487 & 2.345 & 2.509 & 2.335 & 2.531 & 2.347 & 2.552 & 2.358 & 2.572 & 2.388 & 2.592 & 2.416 \\
--1.6 & 2.461 & 2.405 & 2.486 & 2.405 & 2.511 & 2.406 & 2.534 & 2.393 & 2.557 & 2.380 & 2.579 & 2.390 & 2.600 & 2.399 & 2.620 & 2.428 & 2.641 & 2.455 \\
--1.5 & 2.510 & 2.460 & 2.536 & 2.461 & 2.561 & 2.462 & 2.585 & 2.446 & 2.609 & 2.431 & 2.631 & 2.438 & 2.653 & 2.444 & 2.673 & 2.472 & 2.693 & 2.498 \\
--1.4 & 2.563 & 2.521 & 2.590 & 2.522 & 2.617 & 2.524 & 2.642 & 2.507 & 2.666 & 2.491 & 2.688 & 2.493 & 2.710 & 2.496 & 2.730 & 2.521 & 2.750 & 2.543 \\
--1.3 & 2.623 & 2.588 & 2.651 & 2.591 & 2.679 & 2.593 & 2.704 & 2.577 & 2.729 & 2.560 & 2.751 & 2.557 & 2.773 & 2.556 & 2.792 & 2.575 & 2.811 & 2.594 \\
--1.2 & 2.687 & 2.662 & 2.717 & 2.666 & 2.747 & 2.670 & 2.773 & 2.656 & 2.799 & 2.640 & 2.820 & 2.632 & 2.842 & 2.626 & 2.860 & 2.638 & 2.877 & 2.651 \\
--1.1 & 2.757 & 2.741 & 2.789 & 2.748 & 2.820 & 2.755 & 2.847 & 2.746 & 2.874 & 2.734 & 2.895 & 2.721 & 2.917 & 2.709 & 2.933 & 2.710 & 2.949 & 2.714 \\
--1.0 & 2.832 & 2.825 & 2.866 & 2.836 & 2.899 & 2.847 & 2.927 & 2.845 & 2.955 & 2.842 & 2.976 & 2.824 & 2.998 & 2.808 & 3.013 & 2.794 & 3.028 & 2.787 \\
--0.9 & 2.911 & 2.913 & 2.946 & 2.929 & 2.981 & 2.945 & 3.011 & 2.951 & 3.040 & 2.958 & 3.062 & 2.942 & 3.085 & 2.925 & 3.099 & 2.894 & 3.113 & 2.873 \\
--0.8 & 2.992 & 3.004 & 3.030 & 3.025 & 3.067 & 3.045 & 3.098 & 3.060 & 3.129 & 3.077 & 3.153 & 3.067 & 3.176 & 3.056 & 3.191 & 3.012 & 3.205 & 2.974 \\
--0.7 & 3.076 & 3.096 & 3.115 & 3.122 & 3.154 & 3.148 & 3.187 & 3.169 & 3.221 & 3.192 & 3.247 & 3.192 & 3.273 & 3.191 & 3.289 & 3.147 & 3.305 & 3.098 \\
--0.6 & 3.161 & 3.190 & 3.202 & 3.220 & 3.242 & 3.251 & 3.278 & 3.277 & 3.314 & 3.305 & 3.343 & 3.314 & 3.372 & 3.323 & 3.391 & 3.293 & 3.410 & 3.249 \\
--0.5 & 3.248 & 3.286 & 3.290 & 3.320 & 3.332 & 3.354 & 3.370 & 3.384 & 3.408 & 3.414 & 3.441 & 3.432 & 3.473 & 3.450 & 3.498 & 3.441 & 3.522 & 3.418 \\
--0.4 & 3.337 & 3.383 & 3.380 & 3.421 & 3.423 & 3.458 & 3.463 & 3.490 & 3.504 & 3.521 & 3.540 & 3.547 & 3.577 & 3.573 & 3.607 & 3.584 & 3.638 & 3.591 \\
--0.3 & 3.429 & 3.484 & 3.472 & 3.524 & 3.515 & 3.563 & 3.558 & 3.596 & 3.600 & 3.628 & 3.641 & 3.660 & 3.681 & 3.692 & 3.719 & 3.721 & 3.756 & 3.753 \\
--0.2 & 3.525 & 3.590 & 3.568 & 3.631 & 3.611 & 3.670 & 3.655 & 3.704 & 3.699 & 3.737 & 3.743 & 3.774 & 3.788 & 3.810 & 3.831 & 3.853 & 3.876 & 3.901 \\
--0.1 & 3.627 & 3.703 & 3.669 & 3.742 & 3.711 & 3.781 & 3.756 & 3.816 & 3.800 & 3.850 & 3.848 & 3.890 & 3.895 & 3.930 & 3.945 & 3.981 & 3.995 & 4.036 \\
 0.0 & 3.735 & 3.821 & 3.775 & 3.858 & 3.815 & 3.895 & 3.860 & 3.932 & 3.905 & 3.967 & 3.954 & 4.010 & 4.004 & 4.052 & 4.058 & 4.105 & 4.113 & 4.161 \\
 0.1 & 3.848 & 3.945 & 3.885 & 3.978 & 3.923 & 4.012 & 3.967 & 4.051 & 4.011 & 4.089 & 4.062 & 4.133 & 4.113 & 4.176 & 4.171 & 4.228 & 4.229 & 4.279 \\
 0.2 & 3.960 & 4.068 & 3.997 & 4.099 & 4.033 & 4.131 & 4.076 & 4.172 & 4.120 & 4.214 & 4.171 & 4.257 & 4.222 & 4.301 & 4.281 & 4.348 & 4.340 & 4.394 \\
 0.3 & 4.064 & 4.180 & 4.102 & 4.213 & 4.139 & 4.246 & 4.183 & 4.289 & 4.226 & 4.332 & 4.277 & 4.376 & 4.328 & 4.420 & 4.388 & 4.465 & 4.447 & 4.507 \\
 0.4 & 4.154 & 4.277 & 4.195 & 4.315 & 4.236 & 4.352 & 4.281 & 4.394 & 4.326 & 4.436 & 4.378 & 4.482 & 4.430 & 4.527 & 4.488 & 4.576 & 4.547 & 4.623 \\
 0.5 & 4.229 & 4.358 & 4.275 & 4.402 & 4.321 & 4.447 & 4.370 & 4.487 & 4.419 & 4.526 & 4.472 & 4.575 & 4.525 & 4.623 & 4.584 & 4.681 & 4.643 & 4.743 \\
\hline
\end{tabular}
\end{center}
\end{table*}
\end{rotate}

%=========================================================================
%= Table 3 ========================================================================
%=========================================================================
\begin{rotate}
\begin{table*}
\tiny
\begin{center}
\caption{\footnotesize Theoretical $u-g$ for Synthetic GCs with Various Ages ($t$) 
Based on the YEPS Model without ($w/o$) and with ($w$) the Inclusion of the HB Prescription 
(The 5th-order Polynomial Fits to the Model Data).\label{tbl-3}}
\begin{tabular}{rcccccccccccccccccc}
\hline
\hline
[Fe/H] & \multicolumn{18}{c}{$u-g$ ($u$ = WFC3/$F336W$, $g$ = ACS/WFC/$F475W$) } \\
\hline
{} &
\multicolumn{2}{c}{t = 10.0 (Gyr)} &
\multicolumn{2}{c}{10.5} &
\multicolumn{2}{c}{11.0} &
\multicolumn{2}{c}{11.5} &
\multicolumn{2}{c}{12.0} &
\multicolumn{2}{c}{12.5} &
\multicolumn{2}{c}{13.0} &
\multicolumn{2}{c}{13.5} &
\multicolumn{2}{c}{14.0} \\
\hline
& $w/o$ & $w$ & $w/o$ & $w$ & $w/o$ & $w$ & $w/o$ & $w$ & $w/o$ & $w$ & $w/o$ & $w$ & $w/o$ & $w $& $w/o$ & $w$ & $w/o$ & $w$ \\
\hline
--2.5 & 1.437 & 1.439 & 1.441 & 1.420 & 1.444 & 1.398 & 1.448 & 1.380 & 1.452 & 1.361 & 1.456 & 1.363 & 1.460 & 1.364 & 1.465 & 1.374 & 1.469 & 1.384 \\
--2.4 & 1.452 & 1.452 & 1.456 & 1.436 & 1.461 & 1.420 & 1.465 & 1.400 & 1.470 & 1.380 & 1.475 & 1.380 & 1.480 & 1.379 & 1.485 & 1.388 & 1.491 & 1.398 \\
--2.3 & 1.467 & 1.465 & 1.472 & 1.453 & 1.478 & 1.442 & 1.483 & 1.421 & 1.488 & 1.400 & 1.494 & 1.397 & 1.500 & 1.394 & 1.506 & 1.403 & 1.513 & 1.412 \\
--2.2 & 1.483 & 1.479 & 1.489 & 1.471 & 1.495 & 1.464 & 1.501 & 1.443 & 1.507 & 1.421 & 1.514 & 1.415 & 1.521 & 1.410 & 1.528 & 1.419 & 1.535 & 1.428 \\
--2.1 & 1.500 & 1.494 & 1.506 & 1.490 & 1.513 & 1.487 & 1.520 & 1.465 & 1.527 & 1.443 & 1.535 & 1.434 & 1.542 & 1.427 & 1.549 & 1.435 & 1.557 & 1.444 \\
--2.0 & 1.517 & 1.511 & 1.525 & 1.509 & 1.532 & 1.509 & 1.540 & 1.488 & 1.548 & 1.466 & 1.556 & 1.455 & 1.564 & 1.445 & 1.572 & 1.453 & 1.580 & 1.461 \\
--1.9 & 1.535 & 1.528 & 1.544 & 1.530 & 1.552 & 1.533 & 1.561 & 1.512 & 1.569 & 1.490 & 1.578 & 1.477 & 1.586 & 1.465 & 1.595 & 1.472 & 1.603 & 1.479 \\
--1.8 & 1.554 & 1.547 & 1.564 & 1.551 & 1.573 & 1.557 & 1.582 & 1.537 & 1.592 & 1.516 & 1.601 & 1.501 & 1.610 & 1.486 & 1.618 & 1.493 & 1.627 & 1.498 \\
--1.7 & 1.574 & 1.567 & 1.585 & 1.574 & 1.595 & 1.581 & 1.605 & 1.563 & 1.615 & 1.543 & 1.625 & 1.527 & 1.634 & 1.510 & 1.643 & 1.515 & 1.652 & 1.519 \\
--1.6 & 1.595 & 1.589 & 1.607 & 1.598 & 1.618 & 1.607 & 1.629 & 1.590 & 1.639 & 1.572 & 1.649 & 1.554 & 1.659 & 1.535 & 1.668 & 1.539 & 1.677 & 1.541 \\
--1.5 & 1.617 & 1.613 & 1.630 & 1.624 & 1.642 & 1.633 & 1.653 & 1.619 & 1.665 & 1.604 & 1.675 & 1.585 & 1.685 & 1.564 & 1.694 & 1.565 & 1.703 & 1.565 \\
--1.4 & 1.641 & 1.639 & 1.654 & 1.651 & 1.667 & 1.660 & 1.679 & 1.649 & 1.691 & 1.637 & 1.702 & 1.617 & 1.713 & 1.596 & 1.722 & 1.594 & 1.731 & 1.591 \\
--1.3 & 1.665 & 1.668 & 1.680 & 1.679 & 1.694 & 1.689 & 1.706 & 1.680 & 1.719 & 1.672 & 1.730 & 1.653 & 1.742 & 1.632 & 1.751 & 1.626 & 1.760 & 1.620 \\
--1.2 & 1.691 & 1.699 & 1.707 & 1.710 & 1.722 & 1.720 & 1.735 & 1.714 & 1.749 & 1.710 & 1.760 & 1.692 & 1.772 & 1.673 & 1.781 & 1.662 & 1.790 & 1.652 \\
--1.1 & 1.719 & 1.733 & 1.735 & 1.743 & 1.751 & 1.752 & 1.766 & 1.751 & 1.780 & 1.750 & 1.792 & 1.735 & 1.804 & 1.718 & 1.813 & 1.702 & 1.822 & 1.688 \\
--1.0 & 1.748 & 1.770 & 1.766 & 1.779 & 1.783 & 1.786 & 1.798 & 1.789 & 1.813 & 1.794 & 1.826 & 1.782 & 1.838 & 1.770 & 1.847 & 1.748 & 1.857 & 1.730 \\
--0.9 & 1.779 & 1.810 & 1.798 & 1.817 & 1.816 & 1.824 & 1.832 & 1.831 & 1.848 & 1.840 & 1.861 & 1.833 & 1.875 & 1.827 & 1.884 & 1.801 & 1.893 & 1.778 \\
--0.8 & 1.813 & 1.853 & 1.832 & 1.858 & 1.852 & 1.865 & 1.869 & 1.876 & 1.886 & 1.890 & 1.900 & 1.888 & 1.914 & 1.890 & 1.924 & 1.861 & 1.933 & 1.834 \\
--0.7 & 1.849 & 1.899 & 1.869 & 1.903 & 1.890 & 1.910 & 1.908 & 1.925 & 1.927 & 1.942 & 1.942 & 1.947 & 1.957 & 1.955 & 1.967 & 1.930 & 1.978 & 1.901 \\
--0.6 & 1.887 & 1.947 & 1.909 & 1.952 & 1.931 & 1.960 & 1.951 & 1.979 & 1.970 & 1.999 & 1.987 & 2.009 & 2.003 & 2.022 & 2.015 & 2.004 & 2.027 & 1.980 \\
--0.5 & 1.929 & 1.997 & 1.952 & 2.006 & 1.975 & 2.016 & 1.997 & 2.038 & 2.018 & 2.059 & 2.036 & 2.073 & 2.054 & 2.091 & 2.068 & 2.083 & 2.082 & 2.071 \\
--0.4 & 1.975 & 2.050 & 2.000 & 2.064 & 2.024 & 2.080 & 2.047 & 2.102 & 2.070 & 2.123 & 2.090 & 2.141 & 2.111 & 2.160 & 2.128 & 2.164 & 2.144 & 2.168 \\
--0.3 & 2.026 & 2.106 & 2.052 & 2.128 & 2.077 & 2.152 & 2.102 & 2.172 & 2.127 & 2.191 & 2.150 & 2.211 & 2.174 & 2.230 & 2.195 & 2.247 & 2.215 & 2.265 \\
--0.2 & 2.083 & 2.166 & 2.110 & 2.198 & 2.136 & 2.230 & 2.163 & 2.247 & 2.190 & 2.263 & 2.216 & 2.284 & 2.243 & 2.303 & 2.269 & 2.330 & 2.295 & 2.359 \\
--0.1 & 2.149 & 2.233 & 2.176 & 2.272 & 2.203 & 2.309 & 2.232 & 2.325 & 2.260 & 2.340 & 2.290 & 2.362 & 2.319 & 2.380 & 2.350 & 2.415 & 2.381 & 2.450 \\
 0.0 & 2.226 & 2.310 & 2.253 & 2.351 & 2.279 & 2.387 & 2.308 & 2.404 & 2.337 & 2.421 & 2.368 & 2.444 & 2.399 & 2.463 & 2.433 & 2.501 & 2.466 & 2.537 \\
 0.1 & 2.313 & 2.401 & 2.338 & 2.433 & 2.363 & 2.462 & 2.391 & 2.484 & 2.419 & 2.505 & 2.450 & 2.530 & 2.481 & 2.553 & 2.515 & 2.588 & 2.549 & 2.622 \\
 0.2 & 2.403 & 2.500 & 2.427 & 2.515 & 2.450 & 2.533 & 2.476 & 2.562 & 2.502 & 2.590 & 2.531 & 2.618 & 2.561 & 2.648 & 2.594 & 2.677 & 2.628 & 2.706 \\
 0.3 & 2.482 & 2.585 & 2.506 & 2.593 & 2.529 & 2.605 & 2.555 & 2.638 & 2.581 & 2.672 & 2.609 & 2.703 & 2.638 & 2.737 & 2.671 & 2.763 & 2.703 & 2.789 \\
 0.4 & 2.544 & 2.647 & 2.570 & 2.666 & 2.596 & 2.682 & 2.624 & 2.715 & 2.652 & 2.747 & 2.682 & 2.778 & 2.712 & 2.811 & 2.746 & 2.840 & 2.780 & 2.870 \\
 0.5 & 2.593 & 2.692 & 2.622 & 2.731 & 2.652 & 2.776 & 2.683 & 2.792 & 2.715 & 2.813 & 2.749 & 2.843 & 2.784 & 2.872 & 2.822 & 2.911 & 2.859 & 2.951 \\
\hline
\end{tabular}
\end{center}
\end{table*}
\end{rotate}

%=========================================================================
%= Table 4 ========================================================================
%=========================================================================
\begin{table*}
\small
\begin{center}
\caption{References to the Observational Data for the Color--Metallicity Relations.\label{tbl-4}}
\begin{tabular}{lccc}
\tableline
\tableline
{Color--Metallicity Relations}  &  Galaxy Name & \multicolumn{2}{c}{References and Selection Criteria} \\
\tableline
			      		 		&  				& Spectroscopic [Fe/H] 				& Broadband Color 							\\
\tableline
The ($g-z$)--[Fe/H] relation		& Milky Way 		& 1, 2 							& 1, 2 									\\
							& M49	 		& 1, 2 							& 1, 2 									\\
							& M87 			& 1, 2 							& 1, 2 									\\
\tableline
The ($u-z$)--[Fe/H] relation  		& Milky Way 		& 3 								& 3 [($U-I$)\tablenotemark{a}, E($B-V$)\,$<$\,0.3]	\\
($u$ = {\it HST}/WFC3 $F336W$)		& NGC 5128		& 4, 5 [$t$\,$>$\,8 Gyr, S/N\,$>$\,10]		& 6 [($U-I$)\tablenotemark{a}]					\\
							& M87  			& 1, 2 							& 7 [($u-z$)\tablenotemark{c}] 						\\
\tableline
The ($u-g$)--[Fe/H] relation		& Milky Way 		& 3 								& 3 [($U-B$)\tablenotemark{b}, E($B-V$)\,$<$\,0.3]	\\
($u$ = {\it HST}/WFC3 $F336W$)		& NGC 5128		& 4, 5 [$t$\,$>$\,8 Gyr, S/N\,$>$\,10]		& 6 [($U-B$)\tablenotemark{b}]					\\
 							& M87  			& 1, 2 							&   7 [($u-g$)\tablenotemark{d}] 						\\
\tableline
\end{tabular}
\tablecomments{\\
\tablenotemark{a} ($u-z$) = 1.160 $\times$ ($U-I$) + 0.434 (see Figures 3 \& 5). \\
\tablenotemark{b} ($u-g$) = 1.296 $\times$ ($U-B$) + 1.412 (see Figures 3 \& 5). \\
\tablenotemark{c} WFC3 $u_{F336W}$ ($u-z$) = 1.143 $\times$ WFPC2 $u_{F336W}$ ($u-z$) $-$ 0.267 (see Figure 3). \\
\tablenotemark{d} WFC3 $u_{F336W}$ ($u-g$) = 1.280 $\times$ WFPC2 $u_{F336W}$ ($u-g$) $-$ 0.344 (see Figure 3). }
\tablerefs{(1) \citet{peng06}; (2) Paper I; (3) Harris (1996, the 2010 edition); (4) \citet{beasley08}; (5) \citet{chung13}; (6) \citet{peng04a,peng04b}; (7) \citet{yoon11a}.}
\end{center}
\end{table*}

%=========================================================================
%= Table 5 ========================================================================
%=========================================================================
\begin{table*}
\tiny
\begin{center}
\caption{The GMM Analysis for the Observed and Modeled Color Distributions in Figure 6. \label{tbl-5}}
\begin{tabular}{ccc ccc ccc c}
\hline
\hline
{Color} & {$\mu_{\rm b}$\tablenotemark{a}} & {$\mu_{\rm r}$\tablenotemark{a}} & {$\sigma_{\rm b}$\tablenotemark{b}} & {$\sigma_{\rm r}$\tablenotemark{b}} & {$f_{\rm b}$\tablenotemark{c}} & {$f_{\rm r}$\tablenotemark{c}} & {$p(\chi^{2})$\tablenotemark{d}} & {$p(DD)$\tablenotemark{d}} & {$p(kurt)$\tablenotemark{d}} \\

\hline
\multicolumn{10}{c}{Modeled Histograms ($N_{\rm tot}$ = 10,000) as a Homoscedastic Case} \\
\hline
$g-z$  &  0.977 $\pm$ 0.002  &  1.439 $\pm$ 0.003  &  0.146 $\pm$ 0.002  &  0.146 $\pm$ 0.002  &  0.619 $\pm$ 0.006  &  0.381 $\pm$ 0.006  &  0.010  &  0.010  &  0.010 \\
$u-z$  &  2.767 $\pm$ 0.010  &  3.785 $\pm$ 0.017  &  0.400 $\pm$ 0.007  &  0.400 $\pm$ 0.007  &  0.726 $\pm$ 0.011  &  0.274 $\pm$ 0.011  &  0.010  &  0.010  &  0.010 \\
$u-g$  &  1.798 $\pm$ 0.006  &  2.446 $\pm$ 0.020  &  0.263 $\pm$ 0.004  &  0.263 $\pm$ 0.004  &  0.856 $\pm$ 0.011  &  0.144 $\pm$ 0.011  &  0.010  &  0.010  &  1.000 \\
\hline
\multicolumn{10}{c}{Observed Histograms ($N_{\rm tot}$ = 362) as a Homoscedastic Case} \\
\hline
$g-z$  &  0.963 $\pm$ 0.010  &  1.353 $\pm$ 0.014  &  0.124 $\pm$ 0.006  &  0.124 $\pm$ 0.006  &  0.627 $\pm$ 0.033  &  0.373 $\pm$ 0.033  &  0.001  &  0.001  &  0.001 \\
$u-z$  &  2.571 $\pm$ 0.030  &  3.412 $\pm$ 0.040  &  0.303 $\pm$ 0.017  &  0.303 $\pm$ 0.017  &  0.668 $\pm$ 0.037  &  0.332 $\pm$ 0.037  &  0.001  &  0.070  &  0.011 \\
$u-g$  &  1.638 $\pm$ 0.033  &  2.123 $\pm$ 0.085  &  0.228 $\pm$ 0.022  &  0.228 $\pm$ 0.022  &  0.786 $\pm$ 0.067  &  0.214 $\pm$ 0.067  &  0.001  &  0.420  &  0.772 \\
\hline
\multicolumn{10}{c}{Modeled Histograms ($N_{\rm tot}$ = 10,000) as a Heteroscedastic Case} \\
\hline
$g-z$  &  0.921 $\pm$ 0.002  &  1.337 $\pm$ 0.007  &  0.093 $\pm$ 0.002  &  0.209 $\pm$ 0.004  &  0.443 $\pm$ 0.011  &  0.557 $\pm$ 0.011  &  0.010  &  0.010  &  0.010 \\
$u-z$  &  2.545 $\pm$ 0.006  &  3.366 $\pm$ 0.011  &  0.216 $\pm$ 0.005  &  0.544 $\pm$ 0.005  &  0.390 $\pm$ 0.010  &  0.610 $\pm$ 0.010  &  0.010  &  0.010  &  0.010 \\
$u-g$  &  1.655 $\pm$ 0.009  &  2.071 $\pm$ 0.011  &  0.156 $\pm$ 0.006  &  0.337 $\pm$ 0.004  &  0.430 $\pm$ 0.021  &  0.570 $\pm$ 0.021  &  0.010  &  0.010  &  1.000 \\
\hline
\multicolumn{10}{c}{Observed Histograms ($N_{\rm tot}$ = 362) as a Heteroscedastic Case} \\
\hline
$g-z$  &  0.943 $\pm$ 0.019  &  1.321 $\pm$ 0.039  &  0.108 $\pm$ 0.015  &  0.145 $\pm$ 0.021  &  0.562 $\pm$ 0.079  &  0.438 $\pm$ 0.079  &  0.001  &  0.124  &  0.001 \\
$u-z$  &  2.442 $\pm$ 0.027  &  3.077 $\pm$ 0.096  &  0.161 $\pm$ 0.037  &  0.473 $\pm$ 0.048  &  0.357 $\pm$ 0.089  &  0.643 $\pm$ 0.089  &  0.001  &  0.449  &  0.011 \\
$u-g$  &  1.519 $\pm$ 0.028  &  1.816 $\pm$ 0.076  &  0.096 $\pm$ 0.033  &  0.310 $\pm$ 0.032  &  0.249 $\pm$ 0.135  &  0.751 $\pm$ 0.135  &  0.001  &  0.614  &  0.772 \\
\hline

\end{tabular}
\tablecomments{\\
\tablenotemark{a} The mean colors and their uncertainties for blue (subscript ``b'') and red (subscript ``r'') GCs. \\
\tablenotemark{b} The standard deviation values in colors and their uncertainties for blue and red GCs. \\
\tablenotemark{c} The number fractions of blue and red GCs ($f_{\rm b}=N_{\rm blue}/N_{\rm total}$ and $f_{\rm r}=N_{\rm red}/N_{\rm total}$). \\
\tablenotemark{d} The probabilities of preferring a unimodal distribution over a bimodal distribution ($p$-values) derived based on the likelihood ratio test ($\chi^{2}$), 
on the separation of the means relative to their variances ($DD$), and on the kurtosis of a distribution ($kurt$).
}
\end{center}
\end{table*}

%=========================================================================
%= Table 6 ========================================================================
%=========================================================================
\begin{table*}
\normalsize
\begin{center}
\caption{The Median Photometric Errors in $g-z$, $u-z$, and $u-g$ of Five Magnitude Bins for the M84 GCs 
with $\sigma_u$\,$<$\,0.2 mag. \label{tbl-6}}
\vspace{0.5cm}
\begin{tabular}{c c c c c}
\hline
\hline
Mag bins & Number of GCs & {\,\,\,\,\,\,\,\,\,\,$g-z$ Error} & {\,\,\,\,\,\,\,\,\,\,$u-z$ Error} & {\,\,\,\,\,\,\,\,\,\,$u-g$ Error}  \\
\hline
            $u_0$ $\le$ 23.5  &  18  &  {\,\,\,\,\,\,\,\,\,\,0.019}  &   {\,\,\,\,\,\,\,\,\,\,0.016}   &  {\,\,\,\,\,\,\,\,\,\,0.018} \\
23.5 $<$ $u_0$ $\le$ 24.5  &  64  &     {\,\,\,\,\,\,\,\,\,\,0.026}  &   {\,\,\,\,\,\,\,\,\,\,0.026}   &  {\,\,\,\,\,\,\,\,\,\,0.027} \\
24.5 $<$ $u_0$ $\le$ 25.5  &  114  &     {\,\,\,\,\,\,\,\,\,\,0.034}  &   {\,\,\,\,\,\,\,\,\,\,0.046}   &  {\,\,\,\,\,\,\,\,\,\,0.043} \\
25.5 $<$ $u_0$ $\le$ 26.5  &  123  &     {\,\,\,\,\,\,\,\,\,\,0.058}  &   {\,\,\,\,\,\,\,\,\,\,0.089}   &  {\,\,\,\,\,\,\,\,\,\,0.085} \\
            $u_0$ $>$  26.5                  & 43     &     {\,\,\,\,\,\,\,\,\,\,0.088}  & {\,\,\,\,\,\,\,\,\,\,0.160}  &  {\,\,\,\,\,\,\,\,\,\,0.161}    \\
\hline
Entire Sample & 362   &   {\,\,\,\,\,\,\,\,\,\,0.041}   &  {\,\,\,\,\,\,\,\,\,\,0.058}    &  {\,\,\,\,\,\,\,\,\,\,0.056}  \\
\hline
\end{tabular}
\end{center}
\end{table*}

%=========================================================================
%= Table 7 ========================================================================
%=========================================================================
\begin{table*}
\tiny
\begin{center}
\caption{The GMM Analysis for the Observed and Modeled Color Distributions in Figure 6. \label{tbl-5}}
\begin{tabular}{ccc ccc ccc c}
\hline
\hline
MDF & {$\mu_{\rm mp}$\tablenotemark{a}} & {$\mu_{\rm mr}$\tablenotemark{a}} & {$\sigma_{\rm mp}$\tablenotemark{b}} & {$\sigma_{\rm mr}$\tablenotemark{b}} & {$f_{\rm mp}$\tablenotemark{c}} & {$f_{\rm mr}$\tablenotemark{c}} & {$p(\chi^{2})$} & {$p(DD)$} & {$p(kurt)$} \\

\hline
\multicolumn{10}{c}{Inferred Metallicity Histograms ($N_{\rm tot}$ = 10,000) Using Linear CMRs as a Homoscedastic Case} \\
\hline
${\rm [Fe/H]}_{g-z}$  &  $-1.412$ $\pm$ 0.006  &  $-0.385$ $\pm$ 0.008  &  0.358 $\pm$ 0.004  &  0.358 $\pm$ 0.004  &  0.623 $\pm$ 0.007  &  0.377 $\pm$ 0.007  &  0.010  &  0.010  &  0.010 \\
${\rm [Fe/H]}_{u-z}$  &  $-1.330$ $\pm$ 0.006  &  $-0.493$ $\pm$ 0.010  &  0.331 $\pm$ 0.004  &  0.331 $\pm$ 0.004  &  0.673 $\pm$ 0.008  &  0.327 $\pm$ 0.008  &  0.010  &  0.010  &  0.010 \\
${\rm [Fe/H]}_{u-g}$  &  $-1.359$ $\pm$ 0.009  &  $-0.525$ $\pm$ 0.020  &  0.418 $\pm$ 0.006  &  0.418 $\pm$ 0.006  &  0.784 $\pm$ 0.013 &   0.216 $\pm$ 0.013  &  0.010  &  0.010  &  0.950 \\

\hline
\multicolumn{10}{c}{Inferred Metallicity Histograms ($N_{\rm tot}$ = 10,000) Using Non-linear, Model CMRs as a Homoscedastic Case} \\
\hline
${\rm [Fe/H]}_{g-z}$  &  $-1.036$ $\pm$ 0.006  &  $-1.030$ $\pm$ 0.006  &  0.679 $\pm$ 0.010  &  0.679 $\pm$ 0.010  &  0.019 $\pm$ 0.003  &  0.981 $\pm$ 0.003  &  0.010  &  0.010  &  1.000 \\
${\rm [Fe/H]}_{u-z}$  &  $-1.156$ $\pm$ 0.006  &  $-1.148$ $\pm$ 0.006  &  0.718 $\pm$ 0.009  &  0.718 $\pm$ 0.009  &  0.025 $\pm$ 0.006  &  0.975 $\pm$ 0.006  &  0.010  &  0.010  &  1.000 \\
${\rm [Fe/H]}_{u-g}$  &  $-1.309$ $\pm$ 0.009  &  $-1.299$ $\pm$ 0.009  &  0.918 $\pm$ 0.013  &  0.918 $\pm$ 0.013  &  0.027 $\pm$ 0.004  &  0.973 $\pm$ 0.004 &  0.010  &  0.010  &  1.000 \\
 
\hline
\multicolumn{10}{c}{Inferred Metallicity Histograms ($N_{\rm tot}$ = 10,000) Using Linear CMRs as a Heteroscedastic Case} \\
\hline
${\rm [Fe/H]}_{g-z}$  &  $-1.449$ $\pm$ 0.011  &  $-0.444$ $\pm$ 0.018  &  0.331 $\pm$ 0.007  &  0.391 $\pm$ 0.010  &  0.578 $\pm$ 0.014  &  0.422 $\pm$ 0.014  &  0.010  &  0.010  &  0.010 \\
${\rm [Fe/H]}_{u-z}$  &  $-1.471$ $\pm$ 0.017  &  $-0.863$ $\pm$ 0.066  &  0.203 $\pm$ 0.022  &  0.497 $\pm$ 0.030  &  0.318 $\pm$ 0.064  &  0.682 $\pm$ 0.064  &  0.010  &  0.020  &  0.010 \\
${\rm [Fe/H]}_{u-g}$  &  $-1.405$ $\pm$ 0.081  &  $-0.660$ $\pm$ 0.200  &  0.399 $\pm$ 0.087  &  0.459 $\pm$ 0.047  &  0.697 $\pm$ 0.222  &  0.303 $\pm$ 0.222  &  0.010  &  0.040  &  0.950 \\

\hline
\multicolumn{10}{c}{Inferred Metallicity Histograms ($N_{\rm tot}$ = 10,000) Using Non-linear, Model CMRs as a Heteroscedastic Case} \\
\hline
${\rm [Fe/H]}_{g-z}$  &  $-1.938$ $\pm$ 0.061  &  $-0.877$ $\pm$ 0.009  &  1.027 $\pm$ 0.037  &  0.440 $\pm$ 0.008  &  0.144 $\pm$ 0.013  &  0.856 $\pm$ 0.013  &  0.010  &  0.220  &  1.000 \\
${\rm [Fe/H]}_{u-z}$  &  $-2.106$ $\pm$ 0.048  &  $-0.975$ $\pm$ 0.011  &  0.970 $\pm$ 0.033  &  0.489 $\pm$ 0.007  &  0.153 $\pm$ 0.012  &  0.847 $\pm$ 0.012  &  0.010  &  0.120  &  1.000 \\ 
${\rm [Fe/H]}_{u-g}$  &  $-2.467$ $\pm$ 0.060  &  $-1.040$ $\pm$ 0.013  &  1.229 $\pm$ 0.034  &  0.562 $\pm$ 0.009  &  0.182 $\pm$ 0.012  &  0.818 $\pm$ 0.012  &  0.010 &   0.100  &  1.000 \\
\hline

\end{tabular}
\tablecomments{\\
\tablenotemark{a} The mean [Fe/H]'s and their uncertainties for metal-poor (subscribed by ``mp'') and metal-rich (subscribed by ``mr'') GCs. \\
\tablenotemark{b} The standard deviation values in [Fe/H] and their uncertainties for metal-poor and rich GCs. \\
\tablenotemark{c} The number fractions of metal-poor and rich GCs ($f_{\rm mp}=N_{\rm metal-poor}/N_{\rm total}$ and $f_{\rm mr}=N_{\rm metal-rich}/N_{\rm total}$).}
\end{center}
\end{table*}

%=========================================================================
%= Table 8 ========================================================================
%=========================================================================
\begin{table}
\normalsize
\begin{center}
\caption{The Basic Observational Data and Best-fit Model Parameters for M87 and M84. \label{tbl-8}}
\vspace{0.5cm}
\begin{tabular}{lcc}
%\vspace{1cm}
\tableline
\tableline
            & M87 (NGC 4486) & M84 (NGC 4374) \\
\tableline
Basic Observational Data &  &   \\
\tableline
The Virgo Cluster Catalog\tablenotemark{a} No. 					&  1316  		&  763   \\
ACS Virgo Cluster Survey\tablenotemark{b} ID 						&  2  			& 6  \\
The total luminosity in $B$-band, $B_T$\tablenotemark{b} 			& 9.58 mag  	& 10.26 mag \\
%The distance modulus, ($m-M$)$\tablenotemark{c} 					& 31.18 		& 31.33 \\
%The absolute mag in $B$-band, $M_B$\tablenotemark{c}			& $-21.6$		& -21.07 \\
The galaxy morphological type\tablenotemark{b}              	& E0   &  E1  \\
\tableline
Best-fit Parameters for the GC Color Simulations &  &  \\
\tableline
The mean [Fe/H] of GC systems, $\langle$[Fe/H]$\rangle$ 		&  $-0.5$ dex  &  $-0.9$ dex \\
The dispersion of [Fe/H] distributions, $\sigma$([Fe/H])	  		&  0.6 dex  &  0.6 dex  \\
The age of GC systems, $t$										&  13.9 Gyr  & 13.0  Gyr\\
References & Paper II  & This study \\
\tableline
\end{tabular}
\tablenotetext{a}{Binggeli et al. 1987}
\tablenotetext{b}{C\^ot\'e et al. 2004}
\end{center}
\end{table}

\end{document}